\documentclass[%
superscriptaddress,
groupedaddress,
unsortedaddress,
showpacs,
amsmath,amssymb,
aps,
prb,
final,twocolumn
]{revtex4}
\usepackage[pdftex]{graphicx}
   \graphicspath{{.}}
\usepackage{epstopdf} 
\usepackage{subfig}
\usepackage{verbatim}
\begin{document}

\title{First-Principles Studies of Luminescence in Ce doped Inorganic Scintillators}
\author{A. Canning}
\affiliation{Lawrence Berkeley National Laboratory, 1 Cyclotron Rd., Berkeley, CA 94720, USA. }
\affiliation{Department of Applied Science, UC Davis, Davis, CA 95616, USA.}
\author{A. Chaudhry}
\affiliation{Lawrence Berkeley National Laboratory, 1 Cyclotron Rd., Berkeley, CA 94720, USA. }
\affiliation{Department of Applied Science, UC Davis, Davis, CA 95616, USA.}
\author{ R. Boutchko}
\affiliation{Lawrence Berkeley National Laboratory, 1 Cyclotron Rd., Berkeley, CA 94720, USA. }
\author{ N. Gr{\o}nbech-Jensen}
\affiliation{Lawrence Berkeley National Laboratory, 1 Cyclotron Rd., Berkeley, CA 94720, USA. }
\affiliation{Department of Applied Science, UC Davis, Davis, CA 95616, USA.}

\date{\today}

%%%%%%%%%%%%%%%   \ead{rbuchko@lbl.gov}
\begin{abstract}
Luminescence in Ce doped materials corresponds to a transition from an excited 
state where the lowest  Ce {5\textit d} level is filled (often called the
 (Ce$^{3+}$)$^*$ state)  to the ground state where a single {4\textit f} level is filled.
  We have performed theoretical calculations based on Density Functional Theory 
  to calculate the ground state band structure of Ce-doped materials as well as 
  the (Ce$^{3+}$)$^*$ excited state. The excited state calculations used a 
  constrained occupancy approach by setting the occupation of the Ce {4\textit f} 
  states to zero and allowing the first excited state above them to be filled. 
  These calculations were performed on a set of Ce doped materials that 
  are known from experiment to be scintillators or non-scintillators to relate 
  theoretically calculable parameters to measured scintillator performance. 
  From these studies we developed a set of criteria based on calculated 
  parameters that are necessary characteristics for bright Ce activated scintillators. 
Applying these criteria to about a hundred new materials we developed 
a list of candidate materials for new bright Ce activated scintillators. 
After synthesis in powder form one of these new materials (Ba$_2$YCl$_7$:Ce) was found 
to be a bright scintillator.
This approach, involving first-principles calculations of modest computing 
requirements was designed as a systematic, high-throughput method to 
aid in the discovery of new bright scintillator materials by prioritization 
and down-selection 
on the large number of potential new materials.
\end{abstract}

%%%%%%%%%%%%%\submitto{Medical Physics}
%\vspace{2pc}
%\keywords{ NEED KEYWORDS}
\maketitle % Comment out if separate title page not required

%%%%%%%%%%%%%%%%%%%%%%%%%%%%%%%
%%%%%%%%% SECTION 1 : INTRODUCTION %%%%%%
%%%%%%%%%%%%%%%%%%%%%%%%%%%%%%%
\section{\label{sec:intro}Introduction}
%\hspace{8mm}

Inorganic scintillators are extensively employed as radiation detector materials in 
many fields of applied and fundamental research such as medical imaging, high 
energy physics, oil exploration, astrophysics and nuclear materials detection for 
homeland security and other applications. \cite{rodnyi1997,Eijk:2008do} The 
ideal scintillator for gamma ray detection must have exceptional performance in terms of 
stopping power, luminosity, proportionality, speed, and cost. Recently, trivalent 
lanthanide dopants have received greater attention for fast and bright scintillators. 
In particular, Ce$^{3+}$ is a favored dopant in many scintillators due to its allowed 
optical  {5\textit d}\textendash{4\textit f} transition which is relatively fast ($\sim$ 20-40~ns) and 
it can be doped onto La,Y, Gd and Lu sites of many high density host materials. 
Consequently, some of the brightest known scintillators are Ce-doped such as 
LaBr$_3$:Ce, \cite{Loef:2001az} LuI$_3$:Ce\cite{Shah:2004id} and YI$_3$:Ce.
\cite{glodo2008mixed} However, crystal growth and production costs remain 
challenging for these materials.\cite{Loef:2008tu,Iltis2006359}

First principles calculations provide a useful insight into chemical and 
electronic properties of materials and hence can aid in the search for 
better materials or guide modification of existing 
materials. 
\cite{derenzo1999prospects,Klintenberg:2002yp,lordi2007first,singh2008applications} 
The theoretical work presented in this paper is part of a larger 
project \textquotedblleft \textit{High-throughput discovery of 
improved scintillation materials}\textquotedblright, which aims to 
synthesize and characterize 
new materials in microcrystal form and select candidates for crystal 
growth.\cite{derenzo2008design} The main aim of the theoretical 
studies presented here is to develop a fast method to select candidate 
Ce activated scintillator materials for synthesis as well as complement the 
experimental work through simulations of promising synthesized materials. 
Preliminary results from our studies have been presented 
earlier.\cite{canning2009first} In this paper we give a detailed account of our 
first-principles calculations and extensive results obtained so far using more 
advanced calculations than presented in our previous work.

\begin{figure}[h]
\begin{center}
\includegraphics[trim=0mm 0mm 0mm 0mm,clip,scale=1]{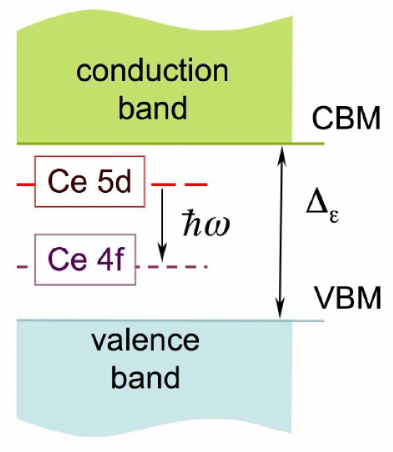}
\end{center}
\caption{
\label{fig:fig1} Schematic diagram of the position of the Ce 
{5\textit d} and {4\textit f} levels relative to the conduction and
valence band of the host material for a Ce activated scintillator. 
VBM is the Valence Band Maximum and CBM is the Conduction Band Minimum.
$\Delta_{\varepsilon}$ is the host material band gap.}
\end{figure}

The basic mechanism for scintillation in a Ce doped material is that an 
incident gamma ray 
will produce a large number of electron-hole (e-h) pairs in the 
host material that  
transfer to the Ce site. The emission of light then corresponds 
to a {5\textit d}\textendash{4\textit f}  
transition on the Ce site from the Ce [Xe]{4\textit f}$^0${5\textit d}$^1$ 
excited state, usually referred to 
as (Ce$^{3+}$)$^*$, to the Ce$^{3+}$ground state [Xe]{4\textit f}$^1${5\textit d}$^0$  
(see Fig.~\ref{fig:fig1}). 
Trapping mechanisms on the host, such as self trapped excitons, 
hole traps or electron 
traps, can quench or reduce the transfer of energy to the Ce site 
(see, for example, 
Ref.~[\onlinecite{rodnyi1997}]  for a more detailed discussion of 
scintillator mechanisms
 and quenching processes).

%%%      FIGURE 1    %%%

Recently, there has been a growing interest in \textit{ab initio} 
calculations of the properties of 
{5\textit d}\textendash{4\textit f} transitions of rare-earth 
ions in solids. Much of this resulted from the 
pioneering work of Dorenbos and collaborators, who in a series of 
papers compiled experimental 
data of this transition and derived semi-empirical models for 
predicting properties of the 
{5\textit d}\textendash{4\textit f} transition and estimated the 
positioning of these states in the host 
gap.\cite{dorenbos20004fn,dorenbos20005d,Dorenbos:2000gs,Dorenbos:2001dg} Most of the 
\textit{ab initio} calculations performed to date for rare-earth 
(RE$^{3+}$) doping use either 
cluster models based on Hartree-Fock or band structure approaches 
based on Density 
Functional Theory (DFT). Usually, in the embedded cluster 
calculations, to reduce computational 
costs, the dopant and the first-shell ions around the dopant are 
allowed to relax while the rest of 
the crystal is kept frozen in the crystalline geometry. This can 
give anomalous 
results\cite{barandiaran2005bond,Pascual:2006mh} and possible 
deficiencies of this local 
relaxation procedure have been discussed recently by Gracia et 
al.\cite{Gracia:2008gs} 
for Ce$^{3+}$ doped YAG (Y$_3$Al$_5$O$_{12}$). Cluster models 
also cannot give the 
positions of the conduction band (CB) and valence band (VB) of 
the host relative to the 
dopant states which is closely related to luminescence 
properties.    

In one of the earliest works using a DFT based approach, Stephan 
et al.\cite{Stephan:2005gc} studied {5\textit 
d}\textendash{4\textit f} transitions for a number of trivalent 
lanthanides using band structure calculations. The effect of 
rare-earth (RE$^{3+}$) doping in semiconducting GaN has also been 
reported.\cite{Mishra:2007ir} However, within the local density 
approximation (LDA) or generalized gradient approximation (GGA) 
to DFT the self-interaction error associated with the localized 
nature of the {4\textit f} shell prohibits the calculation of 
accurate energy differences. There have been attempts to overcome 
this problem using beyond DFT methods, but they have focused on 
studies of bulk Ce compounds.\cite{Petit:2005yn,Loschen:2007dt} 
Recently, Nishida et al.\cite{Nishida:2009bn} studied the 
relationship between the local structure around the Ce$^{3+}$ ion 
and the emission properties of CeF$_3$ and Ce$_2$O$_3$ employing 
a combination of TEM-EELS measurements and first principles band 
structure studies. The {4\textit f}-{5\textit d} energy gap was 
shown to be in qualitative agreement with known experimental 
spectra. 

A band structure approach can also be used to relax the doped 
host matrix to take into account lattice relaxation effects. 
Andriessen et al.\cite{Marsman:2000hf,Andriessen:2005hh} have 
performed such relaxations for a few known Ce$^{3+}$ doped 
scintillating compounds. Recently, they have published detailed 
results for the Stokes shift from lattice relaxation studies of 
{4\textit f}-{5\textit d} excitation of Ce doped lanthanum halide 
scintillators using a band structure approach based on density 
functional theory and ionic cluster calculations using the 
Hartree-Fock method.\cite{Andriessen:2007nd} Watanabe et 
al.\cite{watanabe2008experimental} studied the {4\textit f}-
{5\textit d} absorption spectra of Ce doped LiYF$_4$ using a 
combination of the pseudopotential plane-wave method along with 
the relativistic molecular orbital approach. They found that the 
{4\textit f}-{5\textit d} transitions in the case of Ce$^{3+}$ 
can be attributed to transitions between molecular orbitals since 
Ce$^{3+}$ has a simple [Xe]4\textit{f}$^1$ electronic structure 
implying that Ce {4\textit f}-{5\textit d} transitions can be 
analyzed within the framework of a single-electron approximation. 

Our theoretical calculations for the prediction of candidate 
scintillator materials are based on studies of the Ce {4\textit 
f} and {5\textit d} levels relative to the valence band maximum 
(VBM) and conduction band minimum (CBM) of the host material, 
respectively.\cite{canning2009first} A necessary condition for 
scintillation and luminescence is that the Ce {4\textit f} and 
{5\textit d} levels must be in the gap of the host material. If 
the Ce {4\textit f} level lies in the valence band of the host or 
the {5\textit d} level is in the conduction band there will be no 
Ce activated scintillation or luminescence. If the {5\textit d} 
Ce state lies below but close to the bottom of the conduction 
band then thermal excitation from the {5\textit d} state into the 
conduction band can reduce or quench luminescence. It should also 
be noted that under direct optical excitation of the {4\textit 
f}-{5\textit d} transition some Ce doped systems can show strong 
luminescence but can be weak scintillators due to trapping 
mechanisms on the host that can quench or reduce the transfer of 
energy from the incident gamma ray to the Ce site. 

In the present paper electronic structure calculations of Ce-
doped compounds are performed with the LDA+U (and GGA+U) 
approach.\cite{Anisimov:1997io} This method has been shown in 
previous publications to give a better description of the 
localized {4\textit f} states of Ce compared to LDA or 
GGA.\cite{Loschen:2007dt,Andersson:2007mf} We have tuned the 
empirical U$_{\rm eff}$ parameter for the Ce$^{3+}$ impurity atom 
to match the calculated Ce {4\textit f} to host VBM gap with the 
experimental energy gap for some known scintillating and non-
scintillating Ce-doped compounds. Validation and predictions of 
Ce {4\textit f}\textendash VBM energy gaps in the ground state 
are presented. 

An accurate determination of the Ce {5\textit d}\textendash CBM 
energy gap for the (Ce$^{3+}$)$^*$  state is difficult using 
standard ground state LDA and GGA approximations to density 
functional theory. A ground state calculation with the {4\textit 
f} level filled and the {5\textit d} level empty yields a 
{5\textit d} level that will be higher than when the {5\textit d} 
level is filled and the {4\textit f} level empty. The {4\textit 
f} level is closer to the nuclei than the {5\textit d} level so 
when the {4\textit f} level is emptied the screening effect from 
the positive nuclei will be reduced and the {5\textit d} level 
will move lower. The Stokes shift can also further lower the 
{5\textit d} level but we did not try to model that in our 
simulations. Previous studies have found the Stokes shift to be 
difficult to model accurately with DFT based band structure 
codes.\cite{Andriessen:2007nd} We, therefore, performed excited 
state (constrained LDA) calculations and subsequent analysis to 
allow us to derive a qualitative measure of the {5\textit 
d}\textendash CBM energy gap. Our main aim is not an extremely 
accurate calculation of the {5\textit d} level position, but to 
determine whether or not it is below the CB as this determines if 
luminescence from the Ce site is possible. It should also be 
noted that the host dopant site in our studies is either La, Lu, 
Gd or Y so the CB has {5\textit d} or {4\textit d} character. 
Therefore, systematic errors due to the LDA type treatment of the 
Ce {5\textit d} state will also be present in the determination 
of the CBM yielding, particularly in the case of La,Lu and Gd, a 
reasonably accurate {5\textit d}\textendash CBM separation due to 
cancellation of errors. The size of the supercells in our 
calculations typically prohibited the use of more advanced many-
body methods. Earlier studies of Ce activated scintillators with 
the cluster based Hartree-Fock method found that adding 
configuration interaction only had a minor influence on the 
results. \cite{Andriessen:2007nd} Overall we want to develop a 
high throughput method for the screening of large numbers of new 
materials as candidates for bright Ce activated scintillators so 
we restrict our calculations to computationally fast first 
principles methods that can yield good qualitative results.   

%%%%%%%%%%%%%%%%%%%%%%%%%%%%%%%
%%%%%%%%%% SECTION 2 : CALCULATION %%%%%%
%%%%%%%%%%%%%%%%%%%%%%%%%%%%%%%

\section{Calculation Details }
\label{section:calculation details}
In order to simulate a dopant in a periodic lattice we use the 
supercell approach with periodic boundary conditions. We 
construct a large supercell from periodically repeating the unit 
cell of the host crystal and then replace one of the host 
trivalent site by a Ce atom. We then relax the atomic positions 
while keeping the cell dimensions fixed.  Our basic aim in these 
studies is to model one Ce atom in an infinite host lattice 
however, the supercell approach introduces spurious dopant-dopant 
interactions due to the periodic boundary 
conditions.\cite{Nieminen:2007pv} These interactions can cause a 
broadening of the impurity levels into bands and also 
modification of the valence and conduction band edges which are 
the natural reference energies for the impurity states. We, 
therefore, perform size scaling studies to be sure the supercells 
we use are large enough to produce converged results for the 
properties of interest. 
Once we have relaxed the supercell we perform a ground state 
calculation to determine the position of the Ce {4\textit f} 
level relative to the VBM of the host material. The filled 
{4\textit f} level is typically very localized and atomic in 
nature and has almost no bandwidth so the {4\textit f}\textendash 
VBM gap is well defined.  

To determine if the (Ce$^{3+}$)$^*$ state lies below the CBM we 
perform a constrained LDA (or GGA) calculation by setting the 
occupancy of the Ce {4\textit f} states to zero and filling the 
first state above the {4\textit f} levels. Previous calculations 
for the (Ce$^{3+}$)$^*$ excited state have been performed by 
removing the Ce {4\textit f} states from the basis functions or 
creating a pseudopotential with Ce {4\textit f} states treated as 
core 
states.\cite{Marsman:2000hf,Andriessen:2007nd,Klintenberg:2001vy} 
Our method has the advantage that we can use the same basis set 
and pseudopotential for both excited state and ground state 
calculations allowing direct comparison of energies. We then look 
at the spatial distribution of this excited state to determine if 
it is localized on the Ce or is a delocalized CB character state. 
The level of localization of an electronic state does not have a 
strict mathematical definition, but for the purposes of our 
studies we will define it as the percentage of the normalized 
electron density in a Voronoi cell centered on the Ce atom. We 
will also consider relative localization: a ratio of localization 
of a state on the cerium site to its next largest localization on 
a different cation (La, Lu, Gd or Y).  If the state has no 
localization (the percentage on the Ce atom is very low and the 
ratio is one or below) then we can consider it is a host band 
structure state and is the bottom of the CB. In such a scenario 
any localized state of Ce {5\textit d} character lies above the 
CBM and there is no possibility of scintillation or luminescence. 
If the state is localized on the Ce and has {5\textit d} 
character then we can associate it with the so-called 
(Ce$^{3+}$)$^*$ and a {5\textit d}\textendash{4\textit f} 
transition is possible. We found this procedure for determining 
if there exists a (Ce$^{3+}$)$^*$  state below the CBM necessary, 
as in the systems studied there typically seems to be some level 
of hybridization between the host \textit d character CB and the 
Ce {5\textit d} character states. This will, to some degree, 
delocalize them from an atomic like {5\textit d} state centered 
on the Ce. We, therefore, needed a simple way to characterize the 
lowest \textit d type as a CB or Ce  state without having to 
resort to very large supercell calculations where the electronic 
states and energies were completely converged. We found that for 
wavefunctions localized over many atomic distances the percentage 
in the Voronoi cell as well as the ratio to neighboring cations 
can be low even though to the eye the wavefunction is clearly 
localized in space. Our simple definition of localization does 
not contain any concept of localization distance and poorly 
characterizes states localized over many atomic distances.  We 
will discuss this issue more in the results section where we find 
some of the oxide scintillators like YAP (YAlO$_3$:Ce) have 
(Ce$^{3+}$)$^*$ states localized over many atomic distances.  It 
should also be noted that very localized  (Ce$^{3+}$)$^*$ states 
will tend to have a larger binding energy due to Coulomb 
attraction between the {5\textit d} electron and the nuclei. The 
removal of the {4\textit f} electron reduces the screening of the 
positive nuclei or can be thought of as leaving a hole state 
(compared to the ground state) which has Coulomb attraction with 
the filled {5\textit d} state. The excited state will then have 
more chance of being lower in energy than native exciton states 
on the host which might otherwise reduce or quench the 
scintillation. The Ce {5\textit d} character state lying below 
the CBM is a necessary, but not sufficient, condition for 
scintillation. Related to this, one of the goals of our work is 
study how the Ce {5\textit d} state properties are related to 
scintillation as well as luminescence properties in Ce doped 
materials.

%%%%%% SUBSECTION 2.A : ATOMIC %%%%%%
\subsection{Atomic relaxation studies}

The initial atomic positions and symmetry information of the host 
crystal were taken from the Inorganic Crystal Structure database 
(ICSD).\cite{BERGERHOFF:1983zi,ICSD} The number of atoms in the 
Ce doped supercells was typically 50-150 depending on the size of 
the host unit cell and how many atoms were required for 
reasonable convergence. The Vienna \textit {ab initio} simulation 
package (VASP)\cite{KRESSE:1993kq,Kresse:1996kf,Kresse:1996cu} 
was used for spin-polarized GGA(PBE)\cite{Perdew:1996oq} and LDA 
calculations. The projector-augmented wavefunction (PAW) 
approach, developed by Bl\"ochl\cite{BLOCHL:1994jv} and adapted 
and implemented in VASP\cite{Kresse:1999un} was used for the 
description of the electronic wavefunctions. Plane waves have 
been included up to an energy cut-off of 500~eV. Integration 
within the Brillouin zone was performed with a $\Gamma$ point 
centered grid of {\textit k}-points. The number of irreducible 
{\textit k}-points was typically chosen to be 4 or 8 depending on 
the size and geometry of the supercell. The energy convergence 
criterion was set to 10$^{-6}$~eV and the maximum component of 
force acting on any atom for relaxation of the atomic positions 
after doping with cerium was checked to be less than 0.01~eV$/ 
{\mathrm \AA}$ in every direction. Cerium pseudopotential was 
chosen to include ({5\textit s},{5\textit p},{6\textit 
s},{4\textit f},{5\textit d}) as valence electrons. We have used 
the rotationally invariant method of Dudarev\cite{Dudarev:1998xi} 
as implemented in VASP\cite{Rohrbach:2003tu} for an on-site +U 
correction to treat the cerium {4\textit f} electrons with a 
single parameter U$_{\rm eff}$ = U\textendash J, where the 
Hubbard U parameter is the spherically averaged screened Coulomb 
repulsion energy required for adding an extra electron to the Ce 
4\textit f-states and the parameter J adjusts the strength of the 
exchange interaction. We determined U$_{\rm eff}$ empirically by 
adjusting it to correspond to experimental results (see Section 
IV.A). An artifact of DFT-PBE (or LDA) calculations is that 
unoccupied La {4\textit f} states are positioned at the bottom of 
conduction band.\cite{singh2008applications} However, La 
{4\textit f} states lie higher in 
energy\cite{lang1981study,czyyk1994local} so in our calculations 
we push the La {4\textit f} states higher in the energy plot 
using the the LDA+U approach with the U$_{\rm eff}$ parameter 
taken from Ref.~[\onlinecite{okamoto2006lattice}]. Calculations 
for Gd systems were performed with the Gd\_3 (4\textit f states 
in the core) pseudopotential. We checked two test calculations 
with the regular Gd pseudopotential (4\textit f electrons as 
valence) and found the results to be very similar. We do not 
employ spin-orbit coupling in our calculations as in the case of 
the La halides this was found to only 
move the Ce {5\textit d} states by a maximum of about 0.2 
eV\cite{Andriessen:2007nd} which would not change our qualitative 
conclusions and would increase the computational cost.   

For the purposes of comparison and checking the accuracy of the 
pseudopotentials, ground state density of states (DOS) 
calculations were done for a few Ce-doped systems using the full 
potential linear augmented plane wave (FP-LAPW) code 
WIEN2K.\cite{blaha2001} The relaxed atomic positions from the 
VASP code were used as input to the WIEN2K code. The same 
GGA(PBE) functional was used in the two codes. We kept the 
\textit {\textit k}-point grid and energy convergence criteria 
similar to VASP calculations. The number of plane waves was 
restricted to $R_{MT} \times k_{\mathrm max} = 7$. The fully 
localized limit (FLL) form of GGA+U implementation was used 
within the WIEN2K code to treat the Ce {4\textit f} orbital. The 
value of the U$_{\rm eff}$ parameter was kept the same in the two 
calculations. The results were found to be very similar to PAW 
calculations with VASP with the positions of the various bands 
varying by only a few percent between the two codes.

%%%%%% SUBSECTION 2.B :EXCITED %%%%%%
\subsection{Excited state calculations}
\label{subsec:2b}

In order to study the (Ce$^{3+}$)$^*$ state constrained LDA (and 
GGA) calculations were done at the $\Gamma$ point using the VASP 
code. 
The occupation numbers were manually set to empty the Ce 
{4\textit f} states and fill 
the next highest state. The band decomposed charge density was 
subsequently analyzed 
to derive the localization parameters. 

Excited state calculations were also done within the PAW 
framework as implemented in the ABINIT 
code.\cite{gonze2005brief,gonze2009abinit,Torrent:2008ku} The 
electronic wave functions were expanded in plane waves up to a 
kinetic energy cutoff of 60 Hartree. Self-consistency was 
achieved using a {\textit k}-point grid centered at the $\Gamma$ 
point in reciprocal space. The energy tolerance for the charge 
self-consistency convergence was set to $1 \times 10^{-6}$ 
Hartree. Band-decomposed charge density at the lowest energy 
{\textit k}-point was subsequently analyzed to derive the 
localization parameter. ABINIT calculations were, however, 
limited to compounds with elements having reliable PAW data sets.

%%%%%%%%%%%%%%%%%%%%%%%%%%%%%%%
%%%%%%%%%% SECTION 3: CRITERIA %%%%%%%%%
%%%%%%%%%%%%%%%%%%%%%%%%%%%%%%%
\section{Theoretical Criteria for Scintillation and Luminescence}
\label{section:theoretical criteria}
%\hspace{8mm}

Based on present understanding of scintillation physics and our 
previous first-principles studies of known Ce$^{3+}$ scintillators 
(e.g., YAlO$_3$:Ce (YAP), Lu$_2$SiO$_5$:Ce (LSO), LaCl$_3$:Ce, 
LaBr$_3$:Ce, LaI$_3$:Ce, Lu$_2$Si$_2$O$_7$:Ce (LPS) etc.) 
and non Ce-activated scintillators (e.g., Y$_2$O$_3$:Ce, 
La$_2$O$_3$:Ce, LaAlO$_3$:Ce (LAP)) we have developed 
three criteria based on the following theoretically calculable 
parameters to predict candidate materials for bright Ce$^{3+}$ 
activated scintillation.\cite{canning2009first} 
\begin{enumerate}
\item
The size of the host material bandgap.
\item
The energy difference between the VBM of the host and the Ce {4\textit f} level.
\item
The level of localization of the lowest {\textit d} character excited state needed 
to determine if it is a host CB state or a Ce {5\textit d} character state. 
\end{enumerate}

Criterion 1 is related to the fact that the number of electron-
hole pairs produced 
by an incident gamma ray is inversely proportional to the bandgap 
energy 
although the constant of proportionality varies from material to 
material.\cite{lempicki1994fundamental} 
Therefore, the band gap should be as small as possible but must 
be large enough
to accommodate the Ce {4\textit f} and {5\textit d} states.
LDA and GGA are known to underestimate the bandgap, but for the 
purposes of our calculations it does provide trends in families 
as well as comparative results for similar materials. More 
accurate bandgaps can be calculated theoretically by using more 
advanced methods that go beyond LDA but these methods are 
typically more computationally costly. Therefore for the purposes 
of a qualitative prediction of candidate scintillator materials 
we use LDA and GGA calculations of bandgaps. 

Criterion 2 is related to the cases where the energy transfer to 
the Ce site occurs by sequential hole trapping and electron 
trapping on the Ce site. For these cases if the {4\textit 
f}\textendash VBM gap is large there will be a low probability of 
the hole transferring from the host to the Ce site via thermal 
excitation which will reduce scintillation brightness. 

Criterion 3 as discussed in the previous section is how we 
determine if the lowest \textit d character excited state can be 
associated with a Ce {5\textit d} character state or a conduction 
band state of the host material depending on whether the state is 
localized or not.  

A further expected result of our calculation may be estimation of 
the Ce {5\textit d}\textendash CBM gap. This is in fact extremely 
difficult to calculate accurately as in our large supercells 
there are many \textit d character bands associated with the host 
as well as the those associated with Ce. The bands associated 
with Ce localized {5\textit d} states, even with large 
supercells, typically still have some curvature due to finite 
size effects making the {5\textit d}\textendash CBM gap not well 
defined. It is also a difficult process to scan up through the 
lowest \textit d character bands to determine which are 
associated with Ce {5\textit d} states and which are CB character 
states as most bands show some level of hybridization 
particularly at higher energies and presumably close to to the 
CBM. In many cases the excited state calculation is also slow or 
problematic to converge due to the close proximity of the many 
\textit d character bands to the filled \textit d band. This is 
also one of the reasons we have used different codes such as 
ABINIT and VASP for these calculations as we have found that for 
different systems one code may have better convergence properties 
than the other due to the different minimization methods used. 
For some known scintillators and non-scintillators we have 
performed more detailed studies of the character of the different 
\textit d bands and we will present this in future work. In 
particular for some non-scintillators we do find {5\textit d} 
character Ce states within the conduction band although they 
typically have some hybridization with the host \textit d states. 
The problems in calculating a {5\textit d}\textendash CBM gap are 
not shared in determining the 4\textit f\textendash VBM gap as 
the {4\textit f} state is typically extremely localized on the Ce 
atom giving a flat band even with modest sized supercells. The 
higher energy empty {4\textit f} states are also well separated 
from the filled {4\textit f} states giving fast convergence for 
the ground state calculation.

From the point of view of predicting new bright scintillators it 
would be useful to develop more theoretical criteria related to 
trapping processes on the host that can limit energy transfer to 
the Ce site. Unfortunately, it is difficult to use first 
principles calculations to develop criteria related to these host 
processes as the exact nature of the trapping sites on the host 
are often poorly understood from experiment as well as the 
details of the energy transfer mechanisms to the Ce site. For 
example, accurate calculations of deep self-trapped host excitons 
such as those found in LaF$_3$ often require advanced many body 
theories and involve significant lattice relaxation. The 
dynamical nature of the transfer processes of host excitons and 
hole or electron traps to the Ce site is also difficult to model 
from first principles although there has been work done in 
developing empirical models of these 
processes.\cite{Bizarri:2007sd} Overall though, from an 
energetics point of view, we would expect the transfer of energy 
to the Ce site to be most favored, the deeper the (Ce$^{3+}$)$^*$ 
state is within the bandgap of the host material. For example, if 
the (Ce$^{3+}$)$^*$  state is lower in energy than any host STEs, 
it will preferentially form provided there are no large energy 
barriers to transfer processes from the host to the Ce site. As 
with semiconductor dopant states the depth of the (Ce$^{3+}$)$^*$ 
state in the gap of the host material will be related to its 
level of localization.\cite{Baldereschi:1973kp} This is 
particularly true where the character of the CB and dopant state 
are the same which is the case for the systems studied here. 
Hence we expect criterion 3 to also be related to the brightness 
of a Ce activated scintillator.

%%%%%%%%%%%%%%%%%%%%%%%%%%%%%%%
%%%%%%%%%% SECTION 4: RESULTS %%%%%%%%%
%%%%%%%%%%%%%%%%%%%%%%%%%%%%%%%
\section{Results and Discussions}
\label{section:results}
%\hspace{8mm}

In this section we present results of our theoretical studies on 
Ce doped compounds. The discussion is divided into three 
subsections. The first subsection concerns the ground state 
density of states calculations for Ce$^{3+}$ doped compounds 
specifically, the determination of the U$_{\rm eff}$ parameter 
from experimentally measured Ce {4\textit f}\textendash VBM 
energy gaps. Cell size scaling studies were also performed to 
check the dependence of the U$_{\rm eff}$ parameter on the 
simulation supercell size. The second subsection presents results 
of the excited state calculations. Simulation cell size scaling 
studies are also presented in this section. In the last 
subsection we perform calculations for some new materials doped 
with Ce and apply our theoretical criteria for the prediction of 
new bright candidate Ce activated scintillators. One of the new 
scintillators predicted by our calculations was synthesized in 
microcrystal form and confirmed to be relatively bright. We have 
also generated a database of Ce {4\textit f}\textendash VBM 
energies predicted from first-principles calculations of more 
than 100 compounds.

%%%%%% SUBSECTION 4.A :U$_{\rm eff}$  %%%%%%
\subsection{Ground state calculations : Determination of U$_{\rm 
eff}$ parameter}

U$_{\rm eff}$ can be determined in a self-consistent way as 
demonstrated by Cococcioni et al.\cite{Cococcioni:2005ci} for 
CeO$_2$. However, more frequently U$_{\rm eff}$ is chosen in such 
a way as to reproduce with reasonable accuracy an experimentally 
measured quantity like cell volume, bulk modulus, 
etc.\cite{Loschen:2007dt,Andersson:2007mf,Windiks:2008at} We are 
not aware of any prior publication related to LDA+U type 
calculations for Ce-doped insulators of the type used for 
scintillator detectors. Fortunately, experimental measurements of 
Ce {4\textit f}\textendash VBM gap are known for a few 
scintillators so we chose the U$_{\rm eff}$ parameter to closely 
match these known gaps. 

\begin{figure}[h]
\begin{center}
\includegraphics[trim=18mm 0mm 0mm 11mm,clip,scale=0.5]{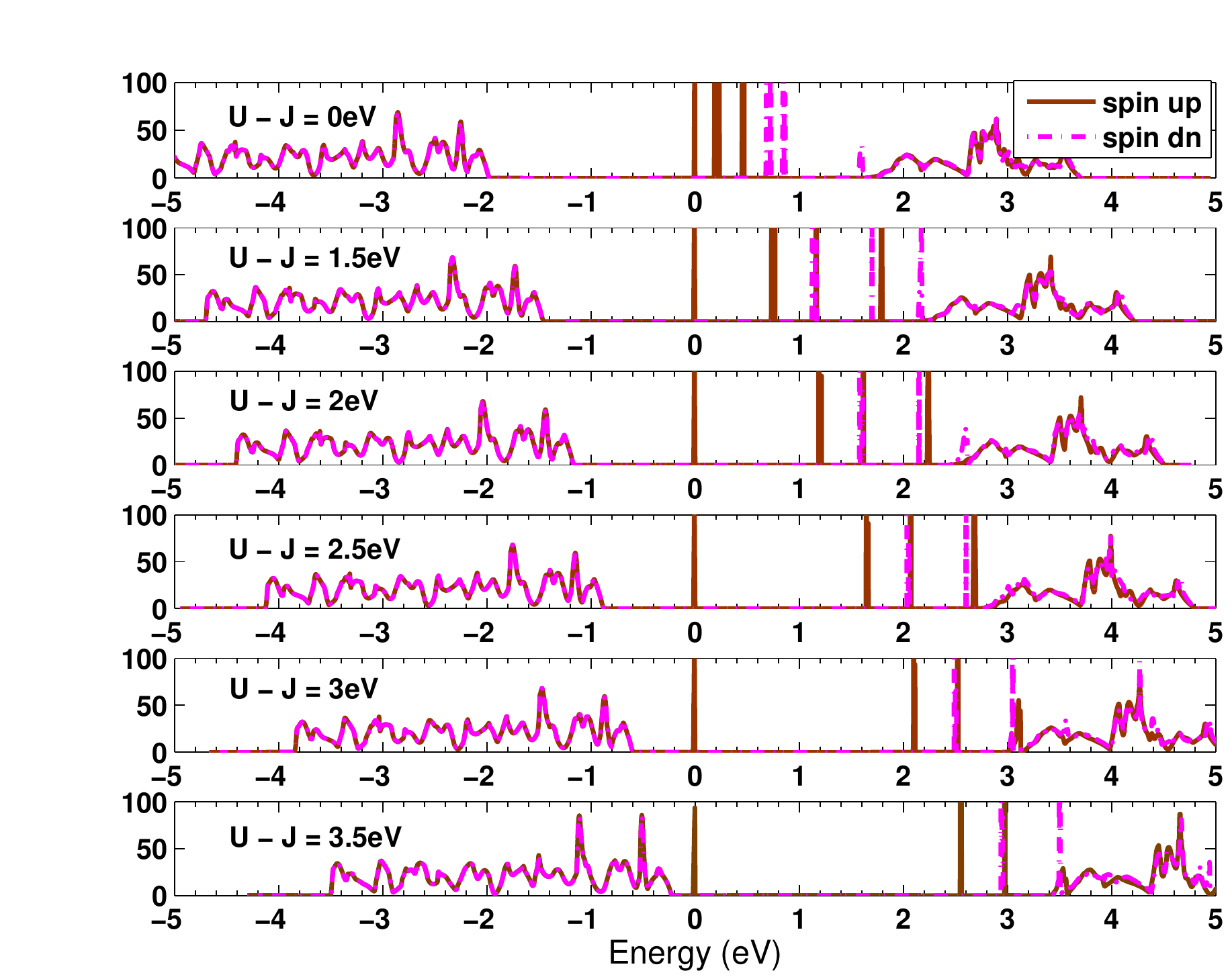}%
\end{center}
\caption{\label{fig:fig2} Ground-state DOS plot for LaBr$_3$:Ce 
from PBE+U spin polarized calculations for different values of 
U$_{\rm eff}$. Fermi level is set at 0. U\textendash J =0~eV 
corresponds to DFT-PBE result. Experimentally estimated 
{4\textit f}\textendash VBM gap is 0.9 $\pm$ 0.4 eV.
\cite{dorenbos2006level} }
\end{figure}

Figure~\ref{fig:fig2} shows the total density of states plot for 
LaBr$_3$:Ce for different values of the U$_{\rm eff}$ parameter 
from a GGA(PBE)+U calculation. The filled Ce {4\textit f} state 
is at the Fermi level which is set to zero. The experimentally 
measured Ce {4\textit f}\textendash VBM gap for LaBr$_3$:Ce is 
0.9~eV ($\pm0.4$~eV).\cite{dorenbos2006level} We observe from the 
figure that the calculated {4\textit f}\textendash VBM energy gap 
using U$_{\rm eff}$ = 2.5~eV matches the experimental data. It is 
important to note that the value of U$_{\rm eff}$ used in the 
literature for bulk Ce(III) compounds (U$_{\rm eff}$ = 4.5~eV for 
PBE functional\cite{Loschen:2007dt,Da-Silva:2007ow}) is different 
from our results. This is mainly due to the itinerant nature of 
{4\textit f} electrons in Ce bulk compounds which participate in 
bonding compared to our doped ionic systems where the single Ce 
{4\textit f} electron is atomic-like in nature. Hence it is 
important to tune the empirical parameter U$_{\rm eff}$ to get a 
close match with experimental data for Ce-doped scintillator 
materials.  It should also be noted in these DOS plots that the 
filled Ce {4\textit f} state is close to being a delta function 
corresponding to a flat band. The {4\textit f}\textendash VBM gap 
is therefore well defined from band structure calculations for 
these types of system.  

%%%      FIGURE 2    %%%

\begin{figure}[h]
\begin{center}
\includegraphics[trim=0mm 0mm 0mm 0mm,clip,scale=0.5]{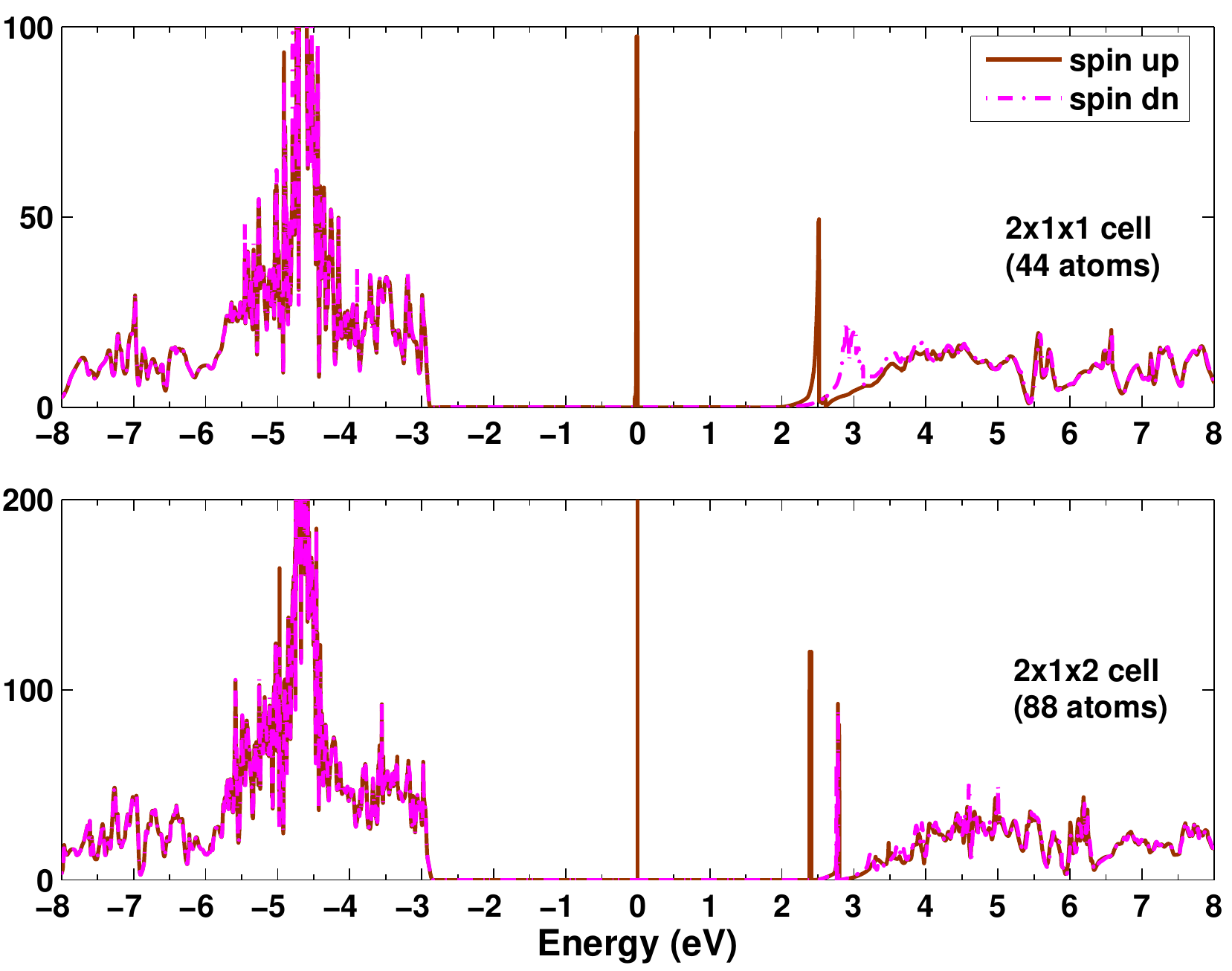}%
\end{center}
\caption{\label{fig:fig3} Density of states (DOS) plot for Lu$_2$Si$_2$O$_7$:Ce 
from PBE+U (U\textendash J = 2.5~eV) spin polarized calculations for two different cell sizes.}
\end{figure}

Figure~\ref{fig:fig3} shows total DOS plots of 
Lu$_2$Si$_2$O$_7$:Ce (LPS) for different simulation cell 
sizes for a fixed U$_{\rm eff}$ parameter. Even for these 
small cell sizes there is negligible variation in the Ce 
{4\textit f}\textendash VBM energy gap ($\sim$2\%) with 
cell size. All the data presented in Table~\ref{tab:table1} 
is for similar or larger cell sizes so we are confident 
that any finite size effects on the Ce {4\textit 
f}\textendash VBM energy gap are below a few percent. 
However, as we show in the next subsection, simulation cell 
size has a greater influence on the localization of the 
excited state.  

%%%      FIGURE 3    %%%

We repeated the calculations for a few systems such as 
YI$_3$:Ce and LaBr$_3$:Ce using the LDA+U functional 
and found negligible change in the results compared to 
GGA(PBE)+U calculations. The choice of the approximation 
to the exchange-correlation functional (PBE or LDA) does 
not affect the position of the impurity (Ce) {4\textit f} 
levels. This is unlike calculations for bulk Ce compounds 
where different values of U$_{\rm eff}$ have been used 
for different functionals.\cite{Da-Silva:2007ow} This is because the 
Ce {4\textit f} atomic-like character changes little for the different functionals.

%%%%%%%%%%%%%%%%%%%%%
%                     Table 1
%%%%%%%%%%%%%%%%%%%%%
\begin{table}[h]%The best place to locate the table environment is directly after its first reference in text
\captionsetup{justification=justified}
\caption{\label{tab:table1}%
Experimentally measured and calculated (PBE+U) 
{4\textit f}\textendash VBM gaps for known Ce-activated
scintillators and non-scintillators.}
\begin{ruledtabular}
\begin{tabular}{p{2.6cm}p{3.3cm}p{2.2cm}}
%\begin{tabular}{lll}
%\hline
{\bf Compound} & {\bf  Measured {4\textit f}\textendash 
VBM gap (eV) }  &  {\bf  PBE+U result (eV)} 
\\
\hline
LaBr$_3$:Ce& 0.9 $\pm$ 0.4&0.9
\\
(scintillator)&(Dorenbos et al.\cite{dorenbos2006level})& 
(U$_{\rm eff}$ = 2.5~eV)
\\[6pt]
Lu$_2$Si$_2$O$_7$:Ce& 2.9 &2.9 
\\
(scintillator)&(Pidol et al. \cite{Pidol:2005bs})&
(U$_{\rm eff}$ = 2.5~eV) 
\\[6pt]
%\hline
Lu$_2$SiO$_5$:Ce  & 3.1   &2.7, 2.9   
\\
(scintillator) & (Joubert et al.\cite{Joubert:2003lg}) & 
(U$_{\rm eff}$ = 2.5~eV) two substitution sites
\\[6pt]
%\hline
YAlO$_3$:Ce &$\sim$ 3.3   &3.0 
\\
(scintillator)&(Nikl et al.\cite{Nikl:2008fr})& (U$_{\rm eff}$ = 2.5~eV)
\\[6pt]
%\hline 
LaI$_3$:Ce   &0.2-0.3   &0.25 
\\
(weak scintillator)    & (Bessiere et al.\cite{Bessiere:2005zc})& 
(U$_{\rm eff}$ = 2.2~eV)
\\[6pt]
%\hline
YPO$_4$:Ce  &$\sim$ 4.0   &3.65 
\\
(weak scintillator)&(Dorenbos \cite{Dorenbos:2004yo})& 
(U$_{\rm eff}$ = 2.5~eV)
\\[6pt]
%\hline
LaAlO$_3$:Ce &$\sim$ 2.0   &2.1 
\\
(non-scintillator)&(van der Kolk et al.\cite{Kolk:2007sd})& 
(U$_{\rm eff}$ = 2.5 eV)
\\[6pt]
%\hline
La$_2$O$_3$:Ce  &$\sim$ 2.8   &2.9 
\\
(non-scintillator)&(Yen et al.\cite{Yen:1996lf})&  
(U$_{\rm eff}$ = 2.5 eV)
\\[6pt]
%\hline
Y$_2$O$_3$:Ce  &$\sim$ 3.4   &3.4 
\\
(non-scintillator)&(Pedrini et al.\cite{Pedrini:2005ol})& 
(U$_{\rm eff}$ = 2.5 eV)
%\\
\end{tabular}
\end{ruledtabular}
\end{table}

Table~\ref{tab:table1} summarizes the results of our 
studies to tune the Ce {4\textit f}\textendash VBM to known 
experimental measurements for Ce doped materials. We see 
from Table~\ref{tab:table1} that GGA(PBE)+U calculations 
with a U$_{\rm eff}$ = 2.5eV give good agreement with 
experimentally measured Ce {4\textit f}\textendash VBM gaps 
for most materials with the exception of LaI$_3$:Ce where 
U$_{\rm eff}$ = 2.2~eV gave the best agreement with 
experiment. LaI$_3$:Ce is one of the smallest bandgap 
scintillator materials so the bonding is more covalent in 
nature than in other scintillator materials. This may 
account for the slightly different character of the Ce 
{4\textit f} state requiring a 0.3~eV lower value of 
U$_{\rm eff}$ than in the other systems. For many of the 
experimental results reported in the table, error bars are 
not quoted in the publications. We have found that in 
scintillator materials the character of the Ce {4\textit f} 
is extremely atomic and very similar for different hosts 
which explains the universality of the U$_{\rm eff}$ value 
in this class of materials. In the case of the heavier host 
La halides there may be some weak dependence of the 
character of the Ce {4\textit f} on the local environment 
surrounding it. From Table~\ref{tab:table1} we can see that 
the {4\textit f}\textendash VBM gap is larger for fluorides 
and progressively decreases as we go down the periodic 
table from F to I. Oxides typically have a larger {4\textit 
f}\textendash VBM gap than sulfides. The U$_{\rm eff}$ 
parameter we found that gives the best fit to experiment 
has no variation from oxides to halides and thus provides a 
potentially simple method to find the Ce {4\textit 
f}\textendash VBM gaps for different types of compounds as 
compared to precise measurements.\cite{Pedrini:2005ol} In 
our calculations reported in subsequent sections for new 
materials we used U$_{\rm eff}$ = 2.5~eV to correct the 
{4\textit f} position except for iodides and sulphides 
where we used 2.2~eV.  

In all the systems studied we found the {4\textit f} level 
to be above the VBM so unlike the {5\textit d} level 
relative to the CBM, the {4\textit f} level position 
relative to the VBM is not a factor in preventing {5\textit 
d}\textendash{4\textit f} emission in Ce doped systems. 

Our studies also revealed that ionic relaxation was 
predominantly influenced by the difference between 
Ce$^{3+}$ ionic radii and the trivalent host cation dopant 
site with the choice of U$_{\rm eff}$ parameter having 
negligible effect. Ce-doped Lu$^{3+}$ and Y$^{3+}$ 
compounds showed significant relaxation as compared to 
La$^{3+}$ compounds primarily because of the almost 10\% 
size mismatch between Ce$^{3+}$ and Lu$^{3+}$ , Y$^{3+}$  
and less than 1\% mismatch between the ionic radii of 
Ce$^{3+}$ and La$^{3+}$ .

%%%%%% SUBSECTION 4.B :Excited state %%%%%%
\subsection{Excited state calculations}

\begin{figure}
\begin{center}
\subfloat[]{\label{fig:4a}\includegraphics[scale=0.5]{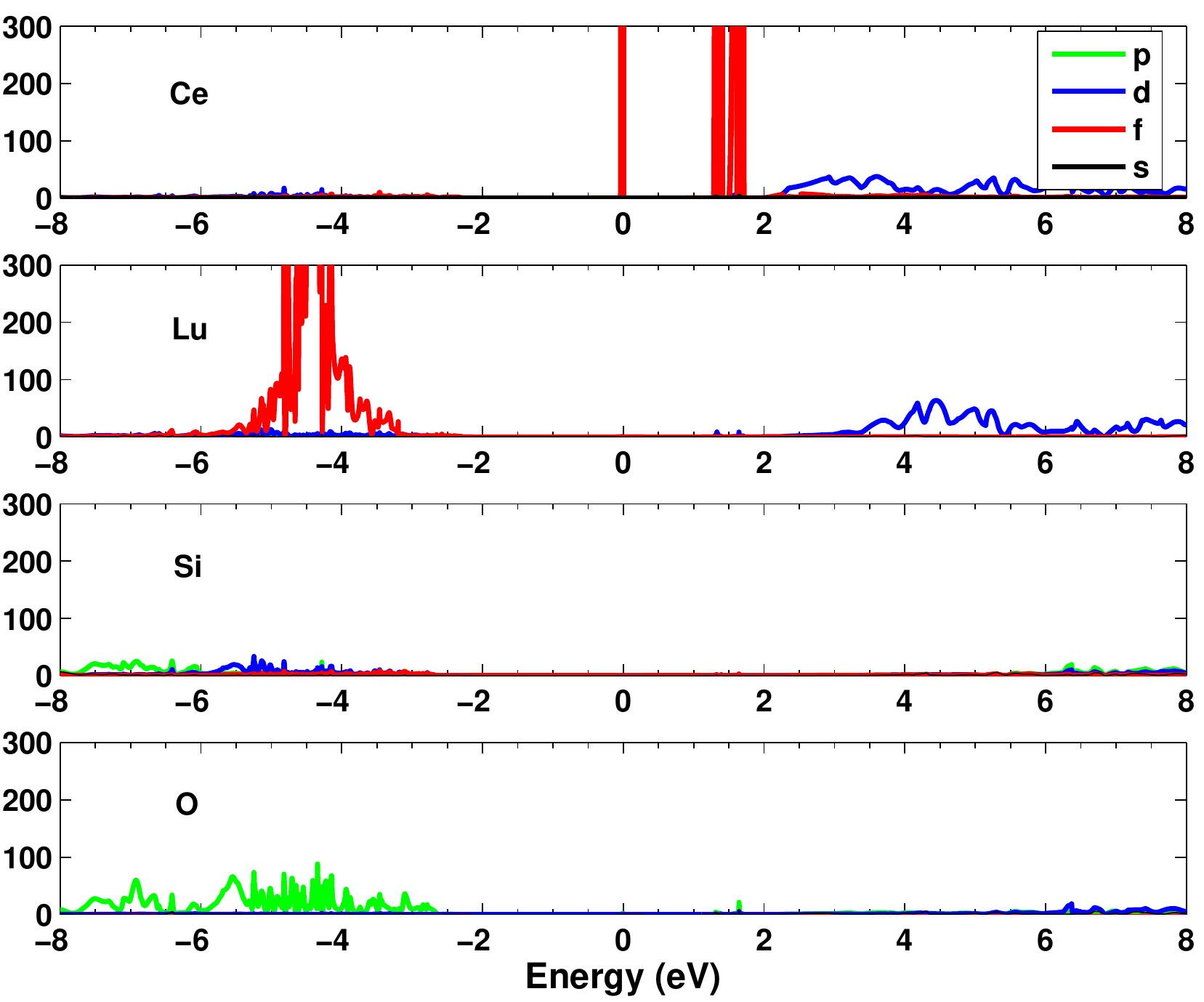}} \qquad
\subfloat[]{\label{fig:4b}\includegraphics[scale=0.5]{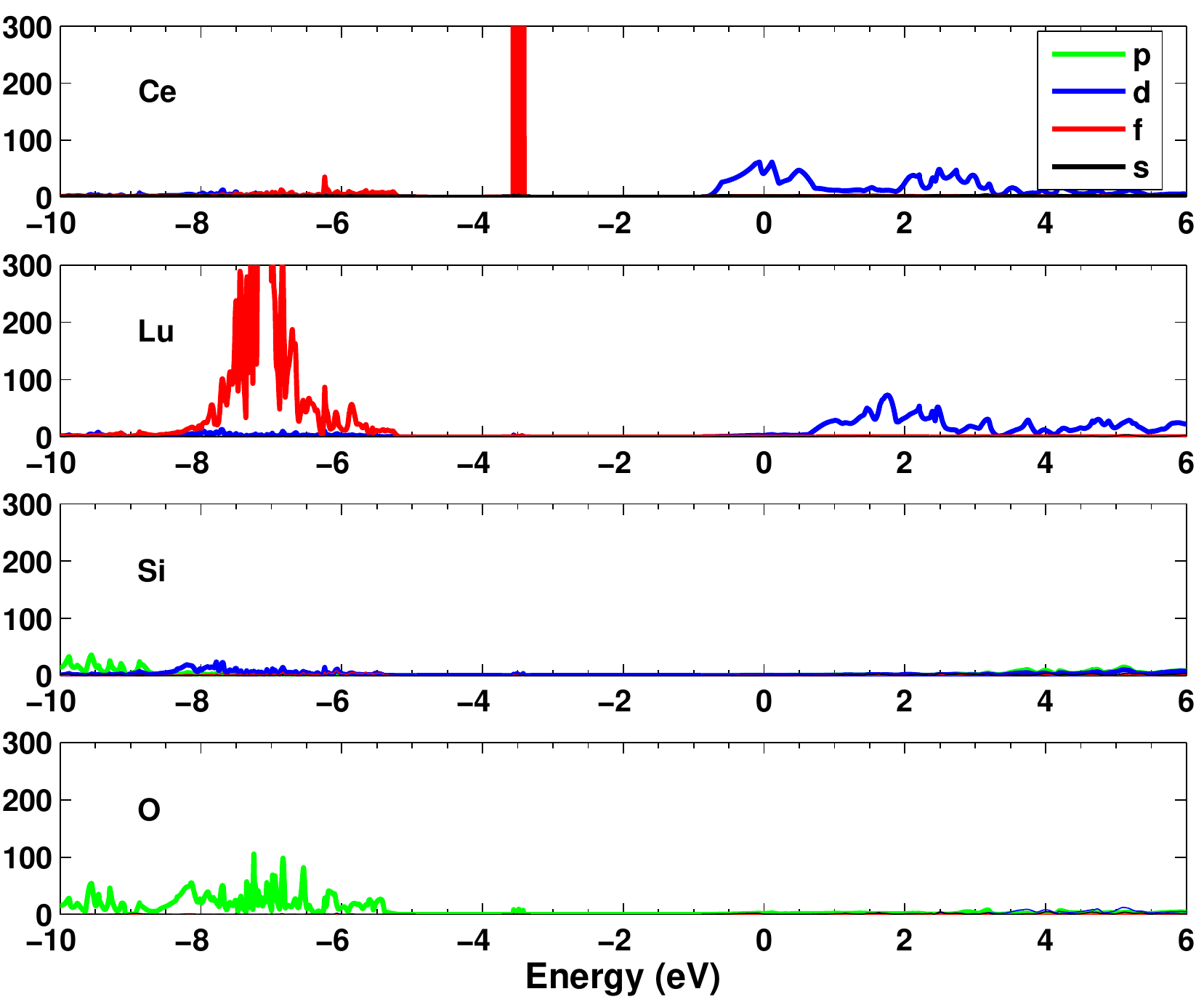}} \\   
\end{center}
\captionsetup{justification=justified}
\caption{\label{fig:fig4} Atom projected partial density of states (DOS) 
plots for GGA(PBE) calculations of Ce doped Lu$_2$Si$_2$O$_7$ 
in the ground state \subref{fig:4a} and excited state \subref{fig:4b}. 
Fermi level is set to 0. \textit f character states are shown in red, 
\textit d states in blue, \textit p characters states in green 
and \textit s character states are shown in black.}
\end{figure}

As described in Section~\ref{subsec:2b}, we performed 
excited state calculations by manually setting the 
occupation of all the Ce {4\textit f} states to zero and 
filling the next highest state. Figure~\ref{fig:fig4}(a-b) 
shows the atom projected partial density of states for 
Lu$_2$Si$_2$O$_7$:Ce (LPS) in the ground state and excited 
state. There is no atomic relaxation in the excited state 
so there is no Stokes shift in our calculations. For this 
system the valence band of the host material consists of O 
\textit p states hybridized with Lu {4\textit f} states and 
the conduction band consists of Lu {5\textit d} character 
states. We can see from the  ground state plot that there 
are Ce {5\textit d} states below the Lu {5\textit d} states 
even in the ground states DOS and these move about 0.5~eV 
lower relative to the CBM in the excited state plot. In the 
excited state plot the excited-fermi level  lies above the 
lowest Ce {5\textit d} level showing the filling of the 
lowest Ce {5\textit d} level. In this system the lowest Ce 
{5\textit d} levels are clearly below the host CB as is 
necessary for the Ce {5\textit d}\textendash{4\textit f} 
transition to occur. As can be seen from this plot the Ce 
{5\textit d} levels have some bandwidth resulting in a 
continuous DOS function for the Ce {5\textit d}  states 
rather than the delta type function we find for the very 
localized Ce {4\textit f} states. The higher energy Ce 
{5\textit d} character states are hybridized with the Lu 
{5\textit d} states. 

%%%      FIGURE 4    %%%

\begin{figure}[h]
\begin{center}
\includegraphics[scale=0.5]{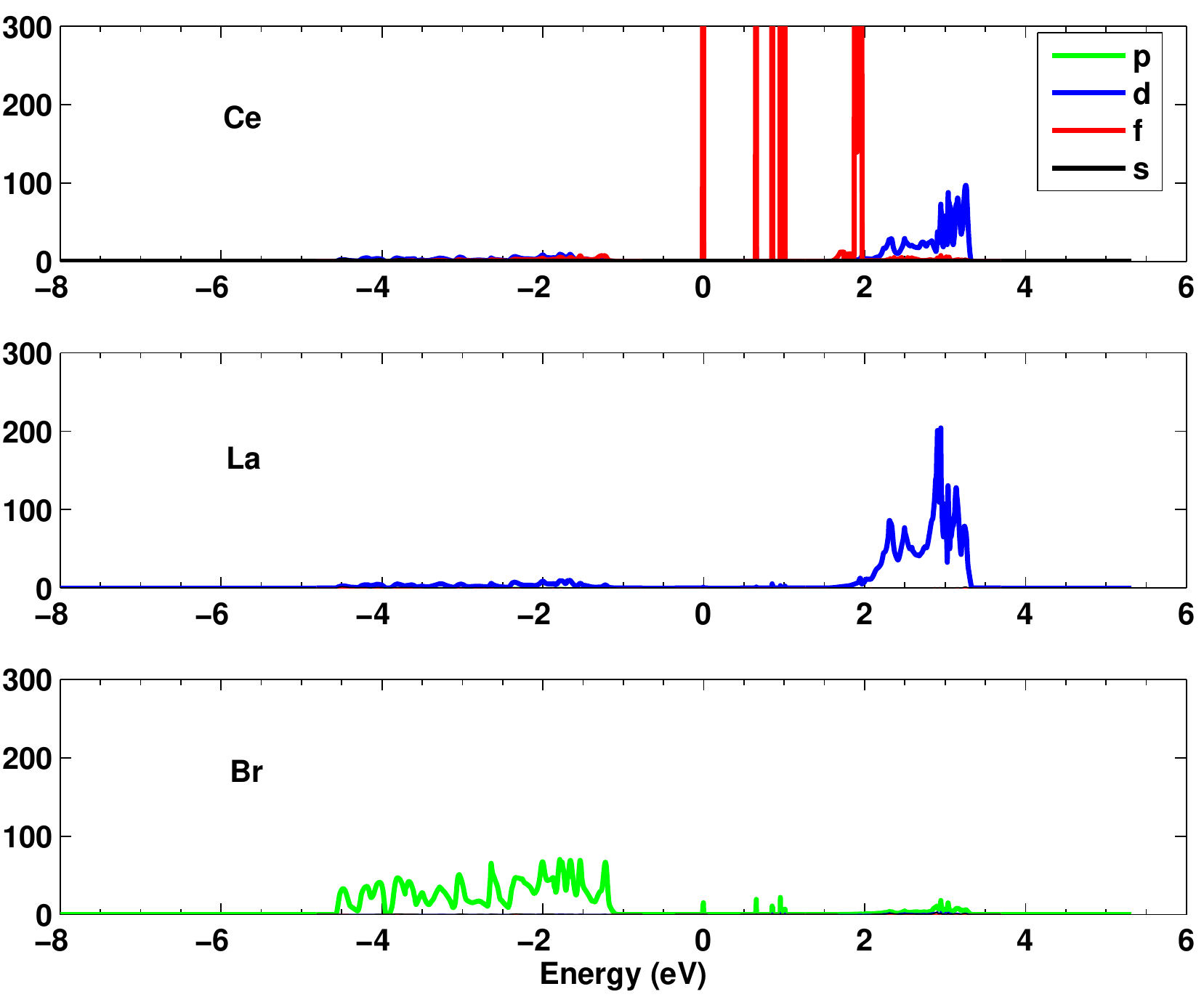}
\end{center}
\captionsetup{justification=justified}
\caption{\label{fig:fig5} Atom projected partial density of states 
plots for LaBr$_3$:Ce in the ground state. Fermi level is set at 0~eV. 
Calculation used the (GGA)PBE functional and Abinit code}
\end{figure}

Figure~\ref{fig:fig5} shows the atom projected partial density 
of states for Ce doped LaBr$_3$ in the ground state which, 
unlike LPS, is more typical of the type of result we obtained 
for different scintillators. The {5\textit d} states on the La and 
Ce are hybridized and occur at the same energy so there 
are no well defined Ce character {5\textit d} states below the 
CB. For these types of systems we find the characterization 
of the lowest filled excited \textit d state in terms of its localization 
on Ce to be the best method 
to determine if it has Ce {5\textit d} character or is a host 
CB character state. 

%%%      FIGURE 5    %%%

\begin{figure*}
%\centering
\subfloat[]{\label{fig:6a}\includegraphics[scale=0.5]{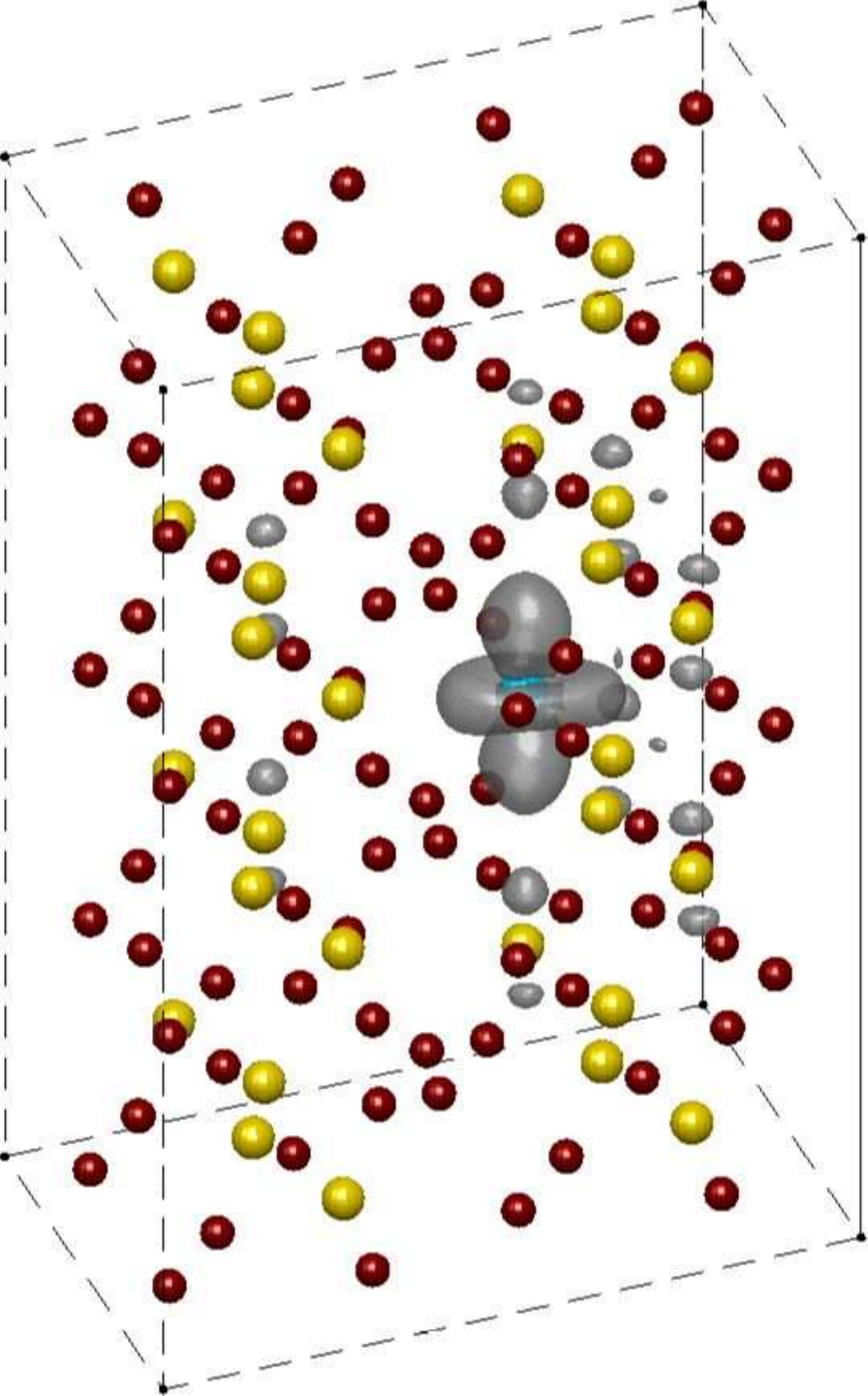}} \qquad
\subfloat[]{\label{fig:6b}\includegraphics[scale=0.45]{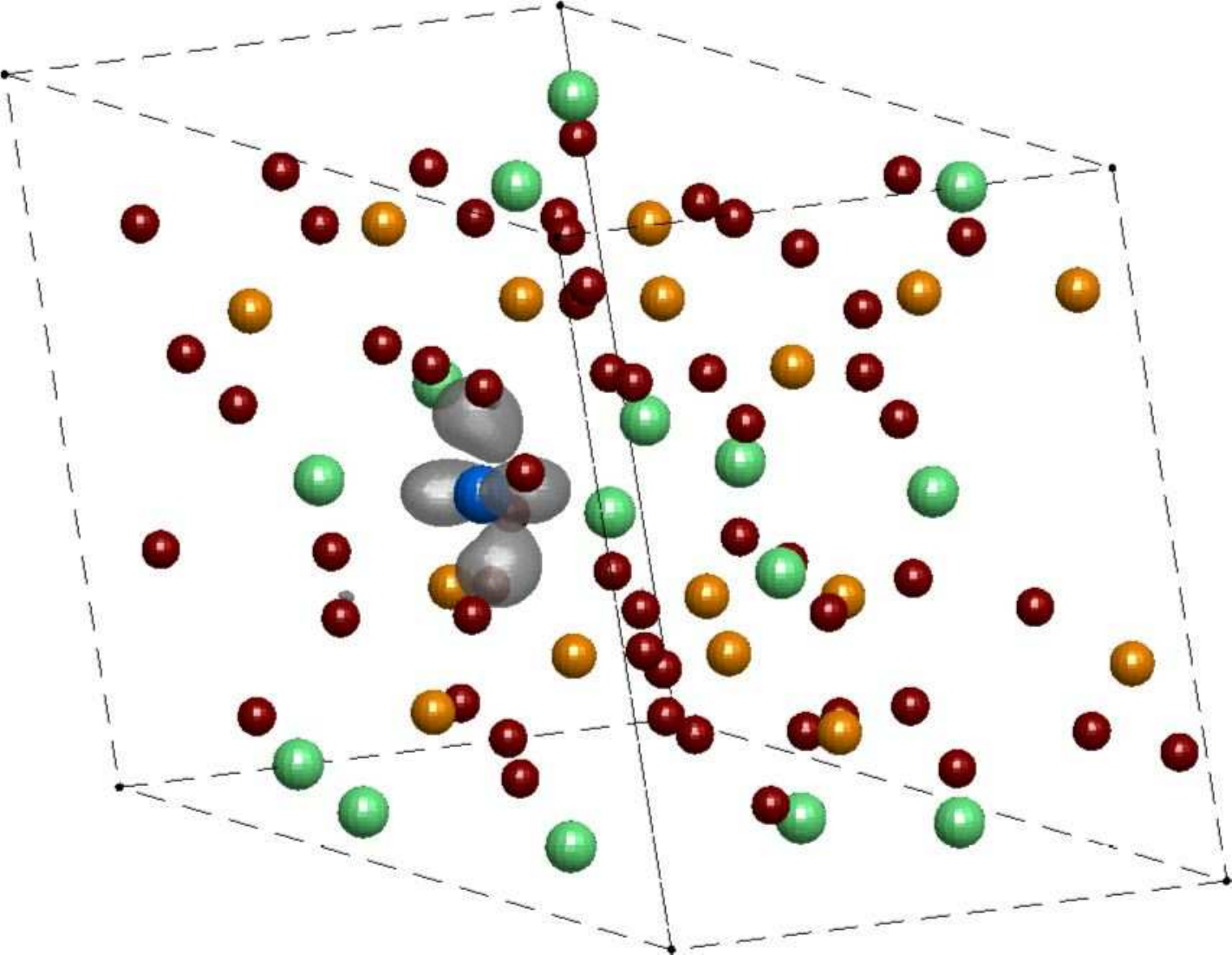}} \qquad   
\subfloat[]{\label{fig:6c}\includegraphics[scale=0.5]{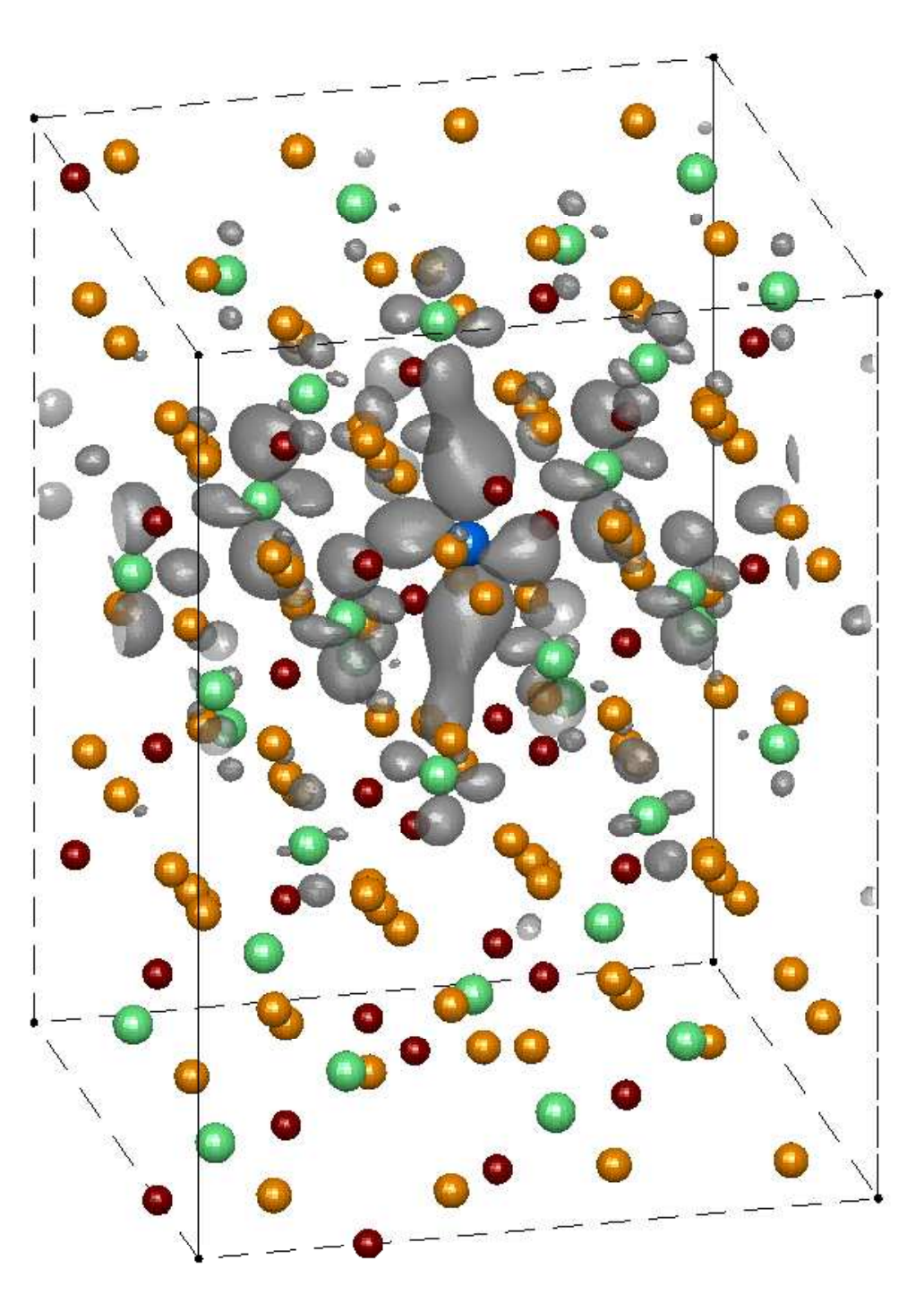}} \\
%\hspace{10pt}
\subfloat[]{\label{fig:6d}\includegraphics[scale=0.5]{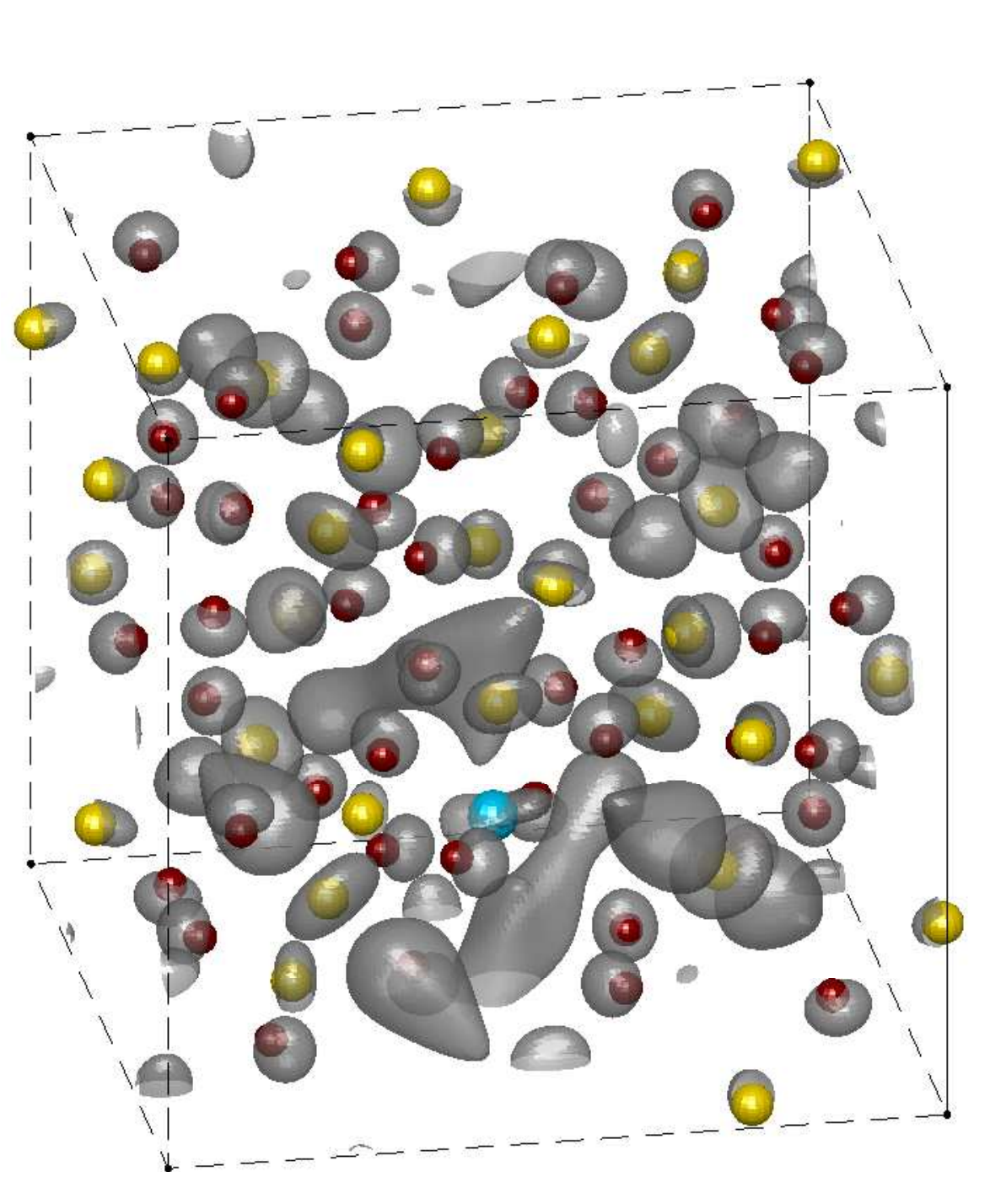}} \qquad
\subfloat[]{\label{fig:6e}\includegraphics[scale=0.4]{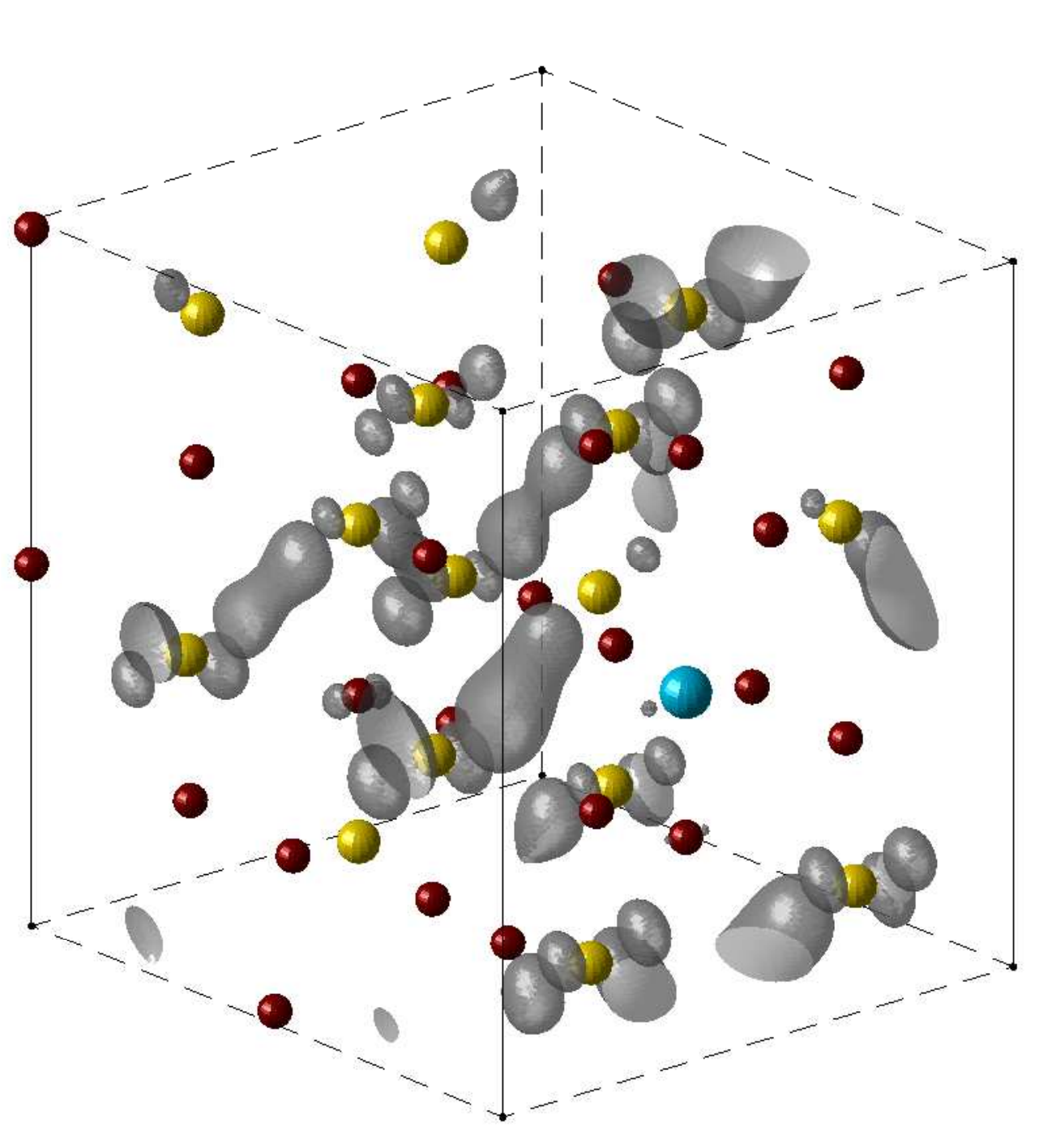}} \qquad
\subfloat[]{\label{fig:6f}\includegraphics[scale=0.66]{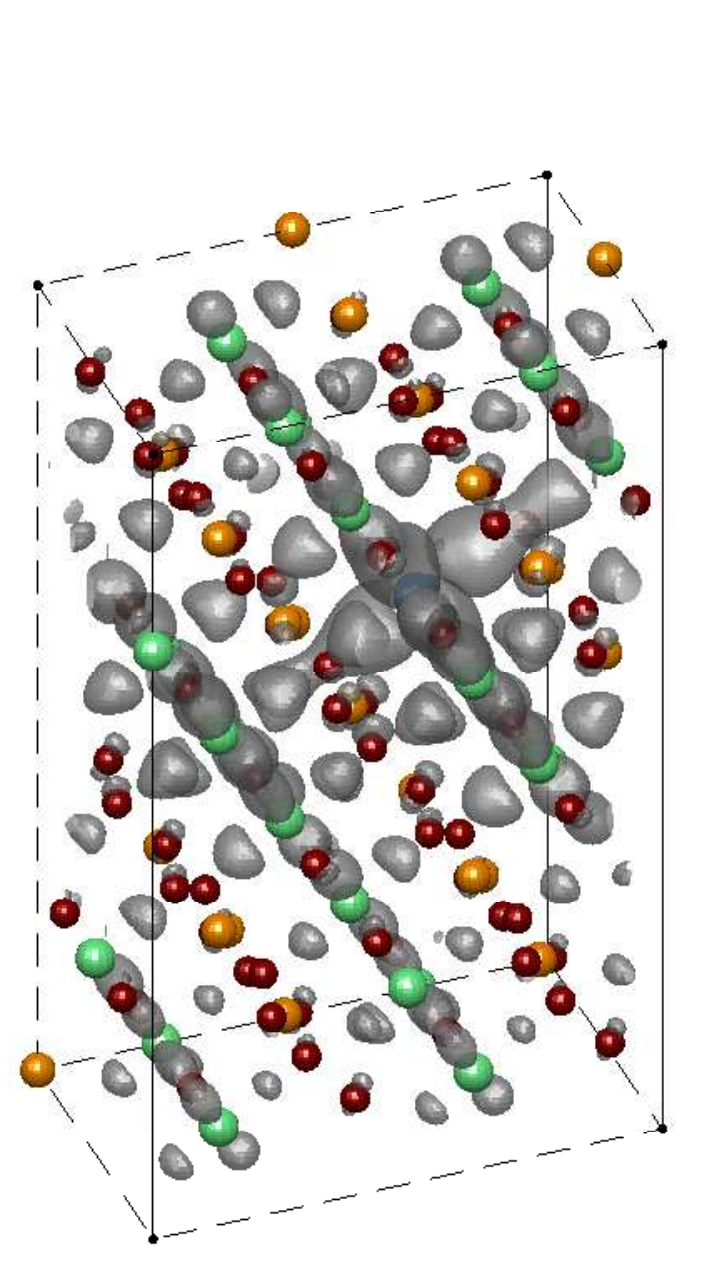}}%
\captionsetup{justification=justified}
\caption{\label{fig:fig6} Lowest {\textit d} character excited state plots 
for Ce scintillators and non-scintillators. Plots show charge density 
isosurfaces of the excited states. Ce atom is shown in blue, 
rare-earth ion (=La, Lu, Y) is in yellow and the anions are shown 
in red. \subref{fig:6a} LaBr$_3$; \subref{fig:6b} Lu$_2$Si$_2$O$_7$; 
\subref{fig:6c} YAlO$_3$; \subref{fig:6d} Y$_2$O$_3$; \subref{fig:6e} 
La$_2$O$_3$; \subref{fig:6f} LaAlO$_3$. The excited state is 
delocalized (very little or no concentration around Ce site) for 
non-scintillating compounds Ð La$_2$O$_3$:Ce, Y$_2$O$_3$:Ce 
and LaAlO$_3$:Ce. However, Ce$^{3+}$ scintillators have good 
localization of the excited state on the Ce site.}
\end{figure*}

Figure~\ref{fig:fig6} shows charge density isosurface plots of the 
first \textit d character excited state at the gamma point for some 
known scintillators and non-scintillators. For the known 
non-scintillators La$_2$O$_3$:Ce, Y$_2$O$_3$:Ce and LaAlO$_3$:Ce 
there is no localization on the Ce and the excited state has a band 
structure character distributed throughout the supercell. On the 
other hand, a localized excited state with \textit d character forms 
on the Ce site for the known scintillators Lu$_2$Si$_2$O$_7$:Ce, 
LaBr$_3$:Ce and YAlO$_3$:Ce. As can be seen from the plots 
there is a large range of localization of the excited state with 
Lu$_2$Si$_2$O$_7$:Ce being much more localized than 
the other systems. 

%%%      FIGURE 6    %%%

Table~\ref{tab:table2} presents a list of our theoretically 
calculated parameters of bandgap, {4\textit f}\textendash 
VBM gap, \% localization and localization ratio for a list 
of known scintillators and non-scintillators compared to 
experimental data for bandgaps and scintillation 
luminosity. As expected LDA consistently underestimates the 
bandgap but does correctly predict the ordering of bandgaps 
for similar materials and families of materials. In all 
these materials the Ce {4\textit f} level is above the VBM 
so the occurrence of the {4\textit f} level within the VB 
never seems to be a factor in quenching luminescence and 
scintillation in Ce doped materials. Also, to the best of 
our knowledge, there is no experimental evidence of Ce 
{4\textit f} states inside the host VB. The main result 
from this table is that we have essentially no localization 
of the lowest excited \textit d state for all the non-
scintillators. The brightest scintillators typically have 
low bandgaps and small {4\textit f}\textendash VBM gaps 
although it should be noted that for scintillation the 
bandgap has to be large enough to accommodate the Ce 
{4\textit f} and {5\textit d} states. Overall, there is 
good qualitative agreement between our three criteria and 
bright scintillators. 

The La halides represent a family of materials that have 
been very heavily studied experimentally for Ce activation 
as they are all scintillators and have a large range of 
bandgaps. LaI$_3$:Ce has a very low band gap of 3.3~eV and 
is thermally quenched at room temperature due to the 
proximity of the excited Ce state to the CBM but has 
reasonable luminosity at 100K.\cite{Bessiere:2005zc} 
Excited state calculations for this system are particularly 
difficult to converge since Ce {5\textit d} states 
hybridize and are very close to the host CB. This is 
consistent with the experimentally estimated {5\textit 
d}\textendash CBM gap of $\sim$ 
0.2~eV.\cite{Bessiere:2005zc} This also leads to a 
relatively low values for the \% localization and ratio. 
The (Ce$^{3+}$)$^*$ excited state is favorably localized 
for LaCl$_3$:Ce and LaBr$_3$:Ce. This agrees with the fact 
that these materials are well known bright scintillators 
used in several gamma ray detection 
applications.\cite{Eijk:2008do} LaBr$_3$:Ce in particular 
has lower bandgap, favorable {4\textit f}\textendash VBM 
gap and reasonable localization on the Ce site. It should 
be noted that the role of host STEs is known to be 
important in the transfer of energy to the Ce site for 
LaCl$_3$:Ce and LaBr$_3$:Ce where the transfer mechanism is 
efficient leading to bright scintillators. In these cases 
the size of the 4{\textit f}\textendash VBM gap will play 
less of a role in determining the 
brightness.\cite{Bizarri:2007sd} LaF$_3$:Ce is an example 
of a system that is known to have a very deep STE of a 
lower energy than the (Ce$^{3+}$)$^*$ excited 
state.\cite{moses1994scintillation}  Even though the lowest 
\textit d character excited state is of Ce {5\textit d} 
character and well localized this limits the transfer of 
energy to the Ce site and results in very low luminosity 
for this particular material. Moreover, the bandgap and 
{4\textit f}\textendash VBM energy gap for this system are 
quite large which leads to comparatively lower e-h pair 
production and a low probability of sequential hole and 
electron capture by Ce. Thus, even if there were no low 
energy host STEs we would not expect this system to be a 
bright scintillator.  

Oxide scintillators in general have wider bandgaps so Ce 
{4\textit f} and {5\textit d} states are mostly better 
separated from the band edges. As we can see from 
Table~\ref{tab:table2} the (Ce$^{3+}$)$^*$ state is 
favorably localized in most of these systems. YAlO$_3$:Ce 
is an example of a system that has a rather low \% 
localization and ratio even though Figure~\ref{fig:fig6} 
clearly shows a localized state. The main reason for this 
is that the state is localized over a few interatomic 
spacings so since our simple measures of localization have 
no measure of localization with distance from the Ce they 
tend to give low values for these types of localized 
states.

%%%%%%%%%%%%%%%%%%%%%
%                        Table 2
%%%%%%%%%%%%%%%%%%%%%
\begin{table*}
\caption{\label{tab:table2}Calculated DFT-PBE bandgaps and 
energy differences for known Ce activated scintillators and 
non-scintillators.  Experimental luminosity data in photons/MeV 
is taken from Ref.~[\onlinecite{%[Comprehensive database of scintillation properties of inorganic materials. ]
scintillatorLBL}] 
and the references therein. ** corresponds to no observed Ce 
emission.}
\begin{ruledtabular}
\begin{tabular}{llcrcr}
 {\bf Compound}&{\bf LDA Bandgap}\footnote{The value in 
 parentheses refers to known experimental 
 bandgaps.}&{\bf Ce {4\textit f}\textendash 
 VBM gap}&\multicolumn{2}{l}{\bf (Ce$^{3+}$)$^{*}$ 
 localization}&{\bf Luminosity}\\
 (atoms in supercell)&(eV)&(eV)& {\bf \% }&{\bf ratio} 
 &(photons/MeV)\\

 \hline
LaF$_3$ (48) & 7.8 (9.7\cite{WIEMHOFER:1990tt}) & 3.5&46 & 9.14 & 2200
\\[6pt]
%\hline
LaCl$_3$ (128) &  4.6 (7\cite{Bizarri:2009cs}) & 1.4 & 40 & 6.08 & 48000
\\[6pt]
%\hline
LaBr$_3$ (128)&3.6 (5.9\cite{dorenbos2006level}) & 0.9 & 21 & 5.70 & 74000
\\[6pt]
%\hline
LaI$_3$ (64)&1.6 (3.3\cite{Bessiere:2005zc})&0.25&
18 % 17.7 
&2.52&200-300\footnote{Luminosity 16000 ph/MeV at 100~K.}
\\[6pt]
%\hline
LaMgB$_5$O$_{10}$ (68)&5.7 (8.8)&2.6&18&2.48&1300
\\[6pt]
%\hline
YI$_3$ (384)&2.8 ($\sim$ 4.13\cite{Srivastava:2010be})&0.6&31&3.48&98600
\\[6pt]
%\hline
YAlO$_3$ (160)&5.4 (8.5-8.9\cite{Basun:1996ni,LUSHCHIK:1994kf})&3.0&21&3.17&21600
\\[6pt]
%\hline
LiGdCl$_4$ (96)&4.6&1.4&74&27.6&64,600
\\[6pt]
%\hline
Lu$_2$Si$_2$O$_7$ (88)&5.5 (7.8\cite{Pidol:2005bs})&2.9&55&6.8&26000
\\[6pt]
%\hline
Lu$_2$SiO$_5$ (64)&4.8 (6.6\cite{Joubert:2003lg})&2.9&33&7.3&33000
\\[6pt]
%\hline
Cs$_2$LiYCl$_6$ (40)&5.0 ($>$ 5.9\cite{Loef:2002mv})&1.8&50&5.8&21600
\\[6pt]
%\hline
$\beta$-KYP$_2$O$_7$ (88)&5.9 ($\sim$7.7\cite{Yuan:2007bn})&2.7&35&6.4&10000
\\[6pt]
%\hline
LaAlO$_3$ (120)&4.0 (5.5\cite{Kolk:2007sd})&2.1&4&1.6&**
\\[6pt]
%\hline
Y$_2$O$_3$ (80)&4.6 (5.8\cite{robertson2000band})&3.4&2&0.63&**
\\[6pt]
%\hline
La$_2$O$_3$ (40)&4.0 (5.3-5.8\cite{Yen:1996lf})&2.9&1&0.15&**
\\[6pt]
%\hline
Lu$_2$O$_3$ (80)&4.7 (5.8)&2.9&2&1.1&**
\\[6pt]
%\hline
Gd$_2$O$_3$ (80)&4.4 (5.4)&2.8&4&0.9&**
\\
\end{tabular}
\end{ruledtabular}
\end{table*}

\begin{figure*}
\centering
\subfloat[]{\label{fig:7a}\includegraphics[scale=0.5]{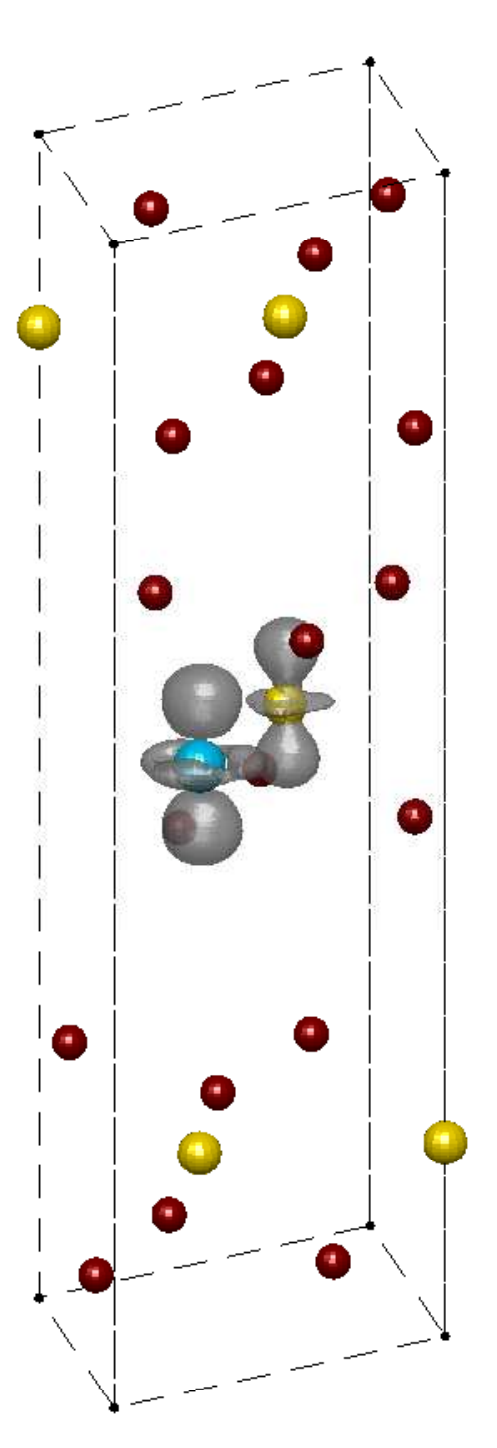}} 
\subfloat[]{\label{fig:7b}\includegraphics[scale=0.55]{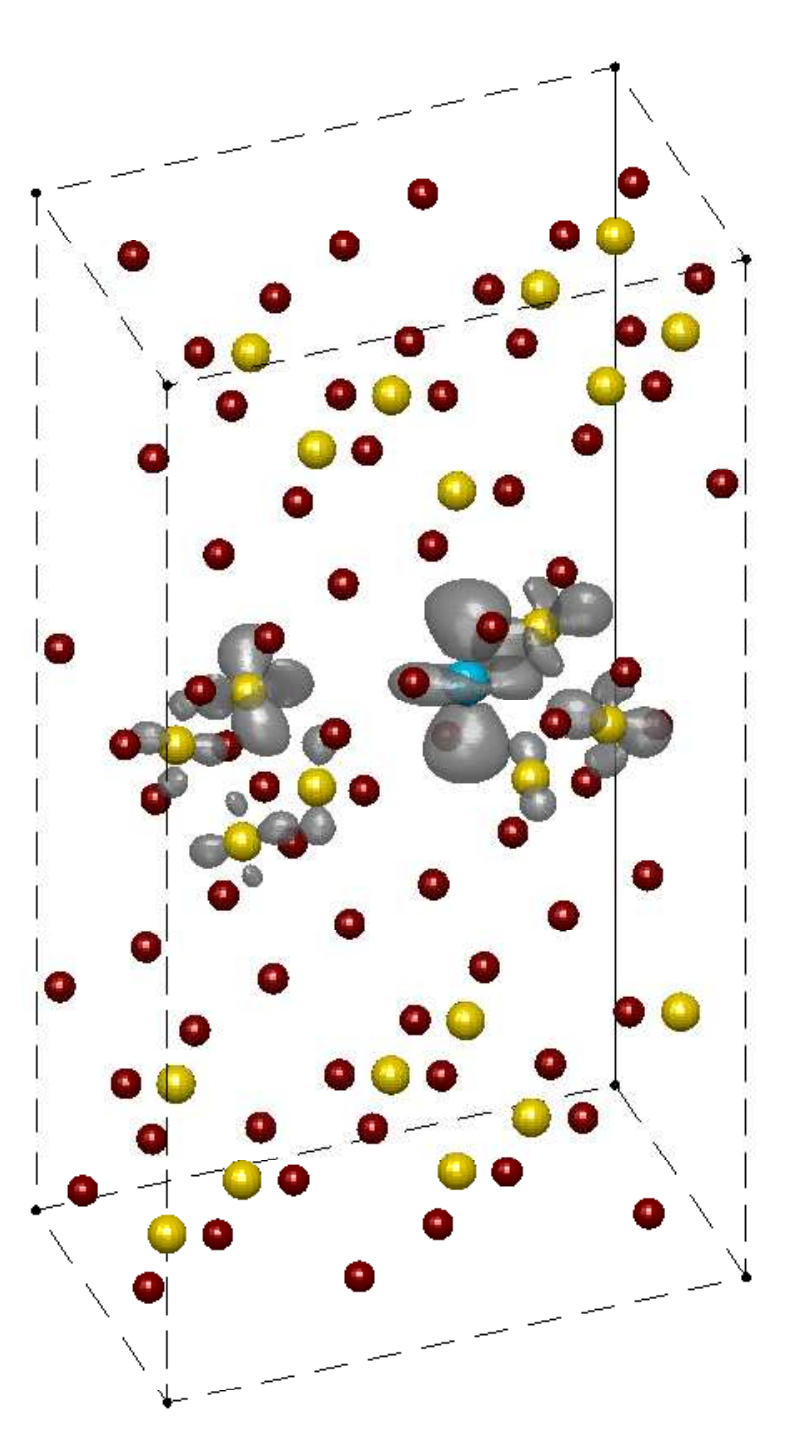}} 
\subfloat[]{\label{fig:7c}\includegraphics[scale=0.75]{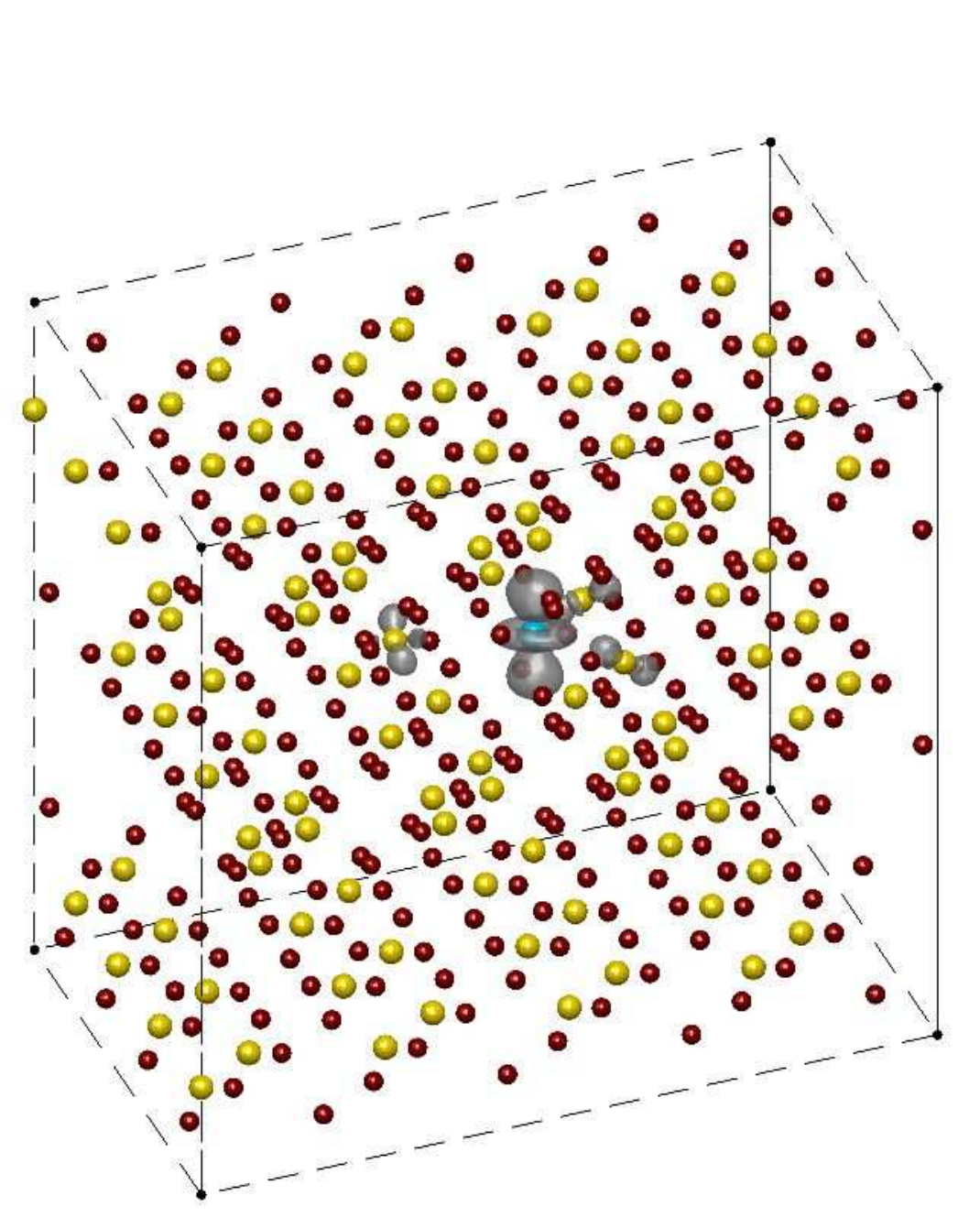}} %
\captionsetup{justification=justified}
\caption{\label{fig:fig7} Excited state charge density isosurface plots of YI$_3$:Ce 
showing the effect of scaling the simulation cell size. All plots are shown at the
 same isosurface threshold. Ce atom is shown in blue, Y is in yellow and Iodine 
 atoms are shown in red. \subref{fig:7a} $1 \times 1 \times 1$ cell (24 atoms); \subref{fig:7b} 
 $2 \times 2 \times 1$ cell (96 atoms); \subref{fig:7c} $4 \times 4 \times 1$ cell (384 atoms). 
 (Ce$^{3+}$)$^*$ excited state localization numbers for these plots are \subref{fig:7a} 
 (40\%, 1.61); \subref{fig:7b} (21\%, 2.67) and \subref{fig:7c} (31.1\%, 3.48).}
\end{figure*}

We also studied the dependence of localization of the excited state with cell size. As an 
example, Figure~\ref{fig:fig7} shows the localization of a bright scintillator YI$_3$:Ce 
with increasing cell size. YI$_3$ has a trigonal crystal structure with \textit a=\textit 
b=7.4864~${\mathrm \AA}$ and \textit c=20.88~${\mathrm \AA}$. In the excited state 
plot for the 24 atom conventional unit cell, even though the dopant Ce$^{3+}$ ion has a 
high percentage localization, the ratio to the next highest Y$^{3+}$ indicates that Ce 
sites in the periodically repeated cells are interacting with each other in the direction of 
the shortest cell dimension (horizontal plane containing the Ce atom and the neighboring 
Y). Now when we scale in the horizontal dimensions for the 96 atom simulation cell we 
find that the Y atoms in the same plane as Ce have some fraction of the excited state, but 
Ce has the highest percentage of the localization of the excited state. This, still, does not 
clearly show a predominating Ce localization expected of a bright scintillator like 
YI$_3$:Ce because the excited state wavefunction is not well localized within the cell 
volume and consequently, there is interaction with the Ce sites in the periodically 
repeated cells in the plane containing lattice vectors \textit a and \textit b. Upon scaling 
the simulation cell size to 384 atoms the excited state becomes predominately 
concentrated on the Ce site. We found that the convergence with cell size varied 
significantly for different host materials. The cell sizes quoted in Table~\ref{tab:table2}  
were chosen to give well converged results for the materials studied and are typically 
smaller than for YI$_3$:Ce. 

%%%      FIGURE 7    %%%

%%%%%% SUBSECTION 4.C :PREDICTIONS %%%%%%
\subsection{Prediction of new candidate Ce scintillators}

The next phase of our theoretical studies was to apply the criteria we have 
developed from studying known scintillators and non-scintillators to the 
prediction of new candidate scintillators. In this study we performed the same 
calculations for about a hundred new materials as we did for the known 
materials. We have chosen new host compounds based on their stopping power and 
their ease of doping with Ce (i.e., the availability of a trivalent sites such 
as Y, La, Lu or Gd for substitution by a Ce atom). In particular, we studied the 
BaY$_m$X$_n$:Ce (X= F,Cl,Br,I) family of materials where BaY$_2$F$_8$:Ce is 
known to be a weak scintillator but the performance of the other materials doped 
with Ce was unknown. The new materials we have studied having the best 
characteristics for bright Ce activation are listed Table~\ref{tab:table3}.

\begin{figure}[h]
\begin{center}
\includegraphics[trim=0mm 0mm 0mm 28mm,clip,scale=0.5]{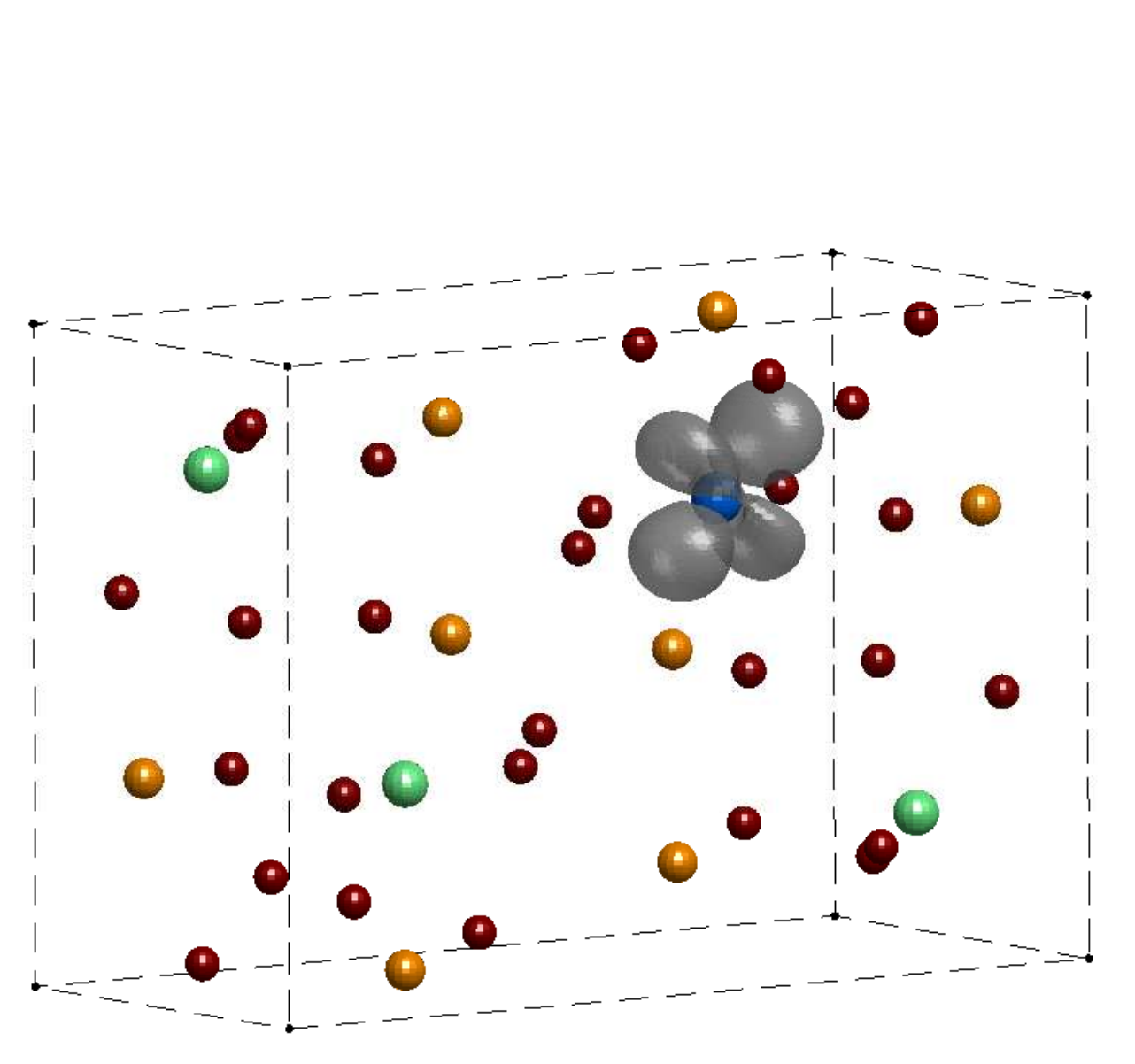}%
\end{center}
\caption{\label{fig:fig8} (Ce$^{3+}$)$^*$ excited state plot for Ba$_2$YCl$_7$:Ce 
at 50\% iso-surface threshold. Ce is shown in blue, Ba in orange, Y is in green 
and the Cl ions are shown in red. }
\end{figure}

We found very strong localization of the (Ce$^{3+}$)$^*$ state for 
Ba$_2$YCl$_7$:Ce (see Fig.~\ref{fig:fig8}). The bandgap and {4\textit 
f}\textendash VBM separation for this material have values that are close to 
those of some of the well known scintillators such as LaCl$_3$:Ce. On the basis 
of our theoretical criteria outlined in Section~\ref{section:theoretical 
criteria}, Ba$_2$YCl$_7$:Ce was expected to be a good candidate for a bright new 
scintillator. It was subsequently synthesized and found to be bright in 
microcrystal form.\cite{Derenzo:2010so} 
In terms of predictions of non-scintillators we have studied many other families 
of materials and have found that for all Y and La host materials containing Ti, 
Zr and Hf there is no localized excited Ce state below the conduction band. 
Ce$^{3+}$ doping in Bi$^{3+}$ host compounds also leads to no localized Ce 
{5\textit d} state below the conduction band. Some of these studies will be the 
subject of future publications.
We have also previously published theoretical work on Ce doped Y and La 
oxyhalides \cite{chaudhry2009first} as well as Y halides 
\cite{boutchko2009cerium} which included known as well as new scintillators.  

%%%      FIGURE 8    %%%

%%%%%%%%%%%%%%%%%%%%%
%                        Table 3
%%%%%%%%%%%%%%%%%%%%%
\begin{table*}
\caption{\label{tab:table3}Calculated bandgaps, {4\textit f}\textendash 
VBM separation and localization for new Ce doped compounds.}
\begin{ruledtabular}
\begin{tabular}{lcclc}
 {\bf Compound}&{\bf LDA Bandgap}&{\bf Ce {4\textit f}\textendash VBM gap}&
 \multicolumn{2}{l}{\bf (Ce$^{3+}$)$^{*}$ localization}\\
 (atoms in supercell)&(eV)&(eV)&{\bf \% }&{\bf ratio}\\
 \hline
CsLa(SO$_4$)$_2$ (48) & 6.0 & 2.0 & 44& 7.5 
\\[6pt]
Ba$_2$YCl$_7$ (40) & 4.7 & 1.6 & 71.7 & 13.2
\\[6pt]
GdIS (96) & 2.5 & 1.3 & 17 & 1.9
\\[6pt]
BaY$_6$Si$_3$B$_6$O$_{24}$F$_2$ (46) & 4.6 & 1.3 & 78 & 15.2
\\[6pt]
Gd$_2$SCl$_4$ (112) & 3.6 & 1.0 & 31.5 & 2.04
\\[6pt]
Cs$_3$Y$_2$Br$_9$ (84) & 3.0 & 1.2 & 34 & 2.69
\\
\end{tabular}
\end{ruledtabular}
\end{table*}

%%%%%%%%%%%%%%%%%%%%%%%%%%%%%%%
%%%%%%%%%% SECTION 5: CONCLUSIONS %%%%%%%%%
%%%%%%%%%%%%%%%%%%%%%%%%%%%%%%%
\section{Conclusions}
\label{section:conclusions}
In this paper we have presented DFT based first principles studies for Ce activated 
scintillator detectors. The main aim of this work was to
determine what theoretically calculable parameters are easily related to luminescence and 
scintillation.  
To more accurately calculate the {4\textit f}\textendash VBM position we used the 
LDA+U approach where we determined U$_{\rm eff}$ by comparison with experimental 
results. We found that a value of U$_{\rm eff}$ of 2.5~eV gave good agreement with 
experiment for a wide range of scintillator materials. Based on this we have calculated 
the {4\textit f}-VBM gap for many known and new materials, some of which are 
presented in Tables~\ref{tab:table2} and~\ref{tab:table3}. We have also generated a 
database of Ce 4f -VBM energies predicted from
first-principles calculations for more than a hundred new compounds.   
We also performed excited state calculations using a constrained LDA approach to 
determine if the first excited \textit d character state was localized on the Ce or was of 
conduction band character. From these studies we developed a set of theoretically 
calculable criteria that characterize bright Ce scintillation. We then validated these 
criteria by studying known scintillators and non-scintillators.  
These criteria were then calculated for about a hundred new materials to determine if they 
were candidates for bright Ce activation. The best candidates are listed in 
Table~\ref{tab:table3}. 
This approach, involving first-principles calculations of modest computing requirements 
was designed as a systematic, high-throughput method to aid the discovery of new bright 
Ce activated scintillator materials. This approach has also been extended to Eu and Pr 
doped systems which will be reported in future publications.  

%%%%%%%%%%%%% ACKNOWLEDGEMENTS %%%%%%%%%%%%%%%%
\begin{acknowledgments}
% \hspace{8mm}
%
We would like to thank Steven Derenzo, Marvin J. Weber, Edith Bourret-Courchesne 
and Gregory Bizarri for many invaluable discussions and constructive criticism. 

The work presented in this paper was supported by the 
U.S. Department of Homeland Security and carried out at the 
Lawrence Berkeley National Laboratory under 
U.S. Department of Energy Contract 
No.  DE-AC02-05CH11231.  

\end{acknowledgments}

%%% BIBLIOGRAPHY - BIBTEX OUTPUT %%%%%%%
%

\end{document}